\RequirePackage{fix-cm}
\documentclass[twocolumn]{svjour3}          
\smartqed  
\usepackage{graphicx}
\journalname{Journal of Superconductivity and Novel Magnetism}

\begin{document}

\title{Topological structures in unconventional scenario for 2D cuprates}


\author{A.S. Moskvin
    \and
    Yu.D. Panov
}


\institute{A.S. Moskvin
\and
Yu.D. Panov
 \at
              Ural Federal University, Ekaterinburg, 
              620083, Russia \\
              \email{alexander.moskvin@urfu.ru}  }

\date{Received: date / Accepted: date}

\maketitle

\begin{abstract}

Numerous experimental data point to cuprates as \emph{d-d} charge transfer unstable systems whose description implies the inclusion of the three many-electron valence states CuO$_4^{7-,6-,5-}$ (nominally Cu$^{1+,2+,3+}$) on an equal footing as a well-defined charge triplet. We introduce a minimal model to describe the charge degree of freedom in  cuprates with the on-site Hilbert space  reduced to only the three states and make use of the S=1 pseudospin formalism. The formalism constitutes a powerful method to study complex phenomena in interacting quantum systems characterized by the coexistence and competition of various ordered states. Overall, such a framework provides a simple and systematic methodology to predict and discover new kinds of orders. In particular, the pseudospin formalism provides the most effective way to describe different topological structures, in particular, due to a possibility of a geometrical two-vector description of the on-site states. We introduce and analyze effective pseudospin Hamiltonian with on-site and inter-site charge correlations, two types of a correlated one-particle transfer and two-particle, or the composite boson transfer.  The latter is of a principal importance for the HTSC perspectives. 
The 2D S=1 pseudospin system is prone to a creation of different topological structures,  which form topologically protected inhomogeneous distributions of the eight local S=1 pseudospin order parameters. We present a short overview of localized topological structures, typical for S=1 (pseudo)spin systems, focusing on unexpected antiphase domain walls in parent cuprates and so-called quadrupole skyrmion, which are believed to be candidates for a topological charge excitation in parent or underdoped  cuprates. Puzzlingly, these unconventional structures can be characterized by an uniform distribution of the mean on-site charge, that makes these invisible for X-rays. Quasiclassical approximation and computer simulation are applied to analyze localized topological defects and evolution of the domain structures  in "negative-$U$" model under charge order-superfluid phase transition.

\keywords{high-T$_c$ cuprates \and charge degree of freedom \and S=1 pseudospin formalism \and topological structures \and unconventional skyrmions}
\end{abstract}

\section{Introduction}
\label{intro}

The origin of high-T$_c$ superconductivity\,\cite{Muller} is presently still a matter of great controversy.  Both copper and novel non-copper based layered high-T$_c$ materials  reveal normal and superconducting state properties very different from that of standard electron-phonon coupled "conventional" \, superconductors.

Copper oxides start out life as insulators in contrast with BCS
superconductors being conventional metals. Unconventional behavior of these materials under charge doping, in particular, a remarkable interplay of charge, lattice, orbital, and spin degrees of freedom, strongly differs from that of ordinary metals and merely resembles that of a doped Mott insulator. In addition to the occurrence of unconventional d-wave superconductivity the phase diagram of the high-T$_c$ cuprates does reveal a flurry of various anomalous electronic properties. In normal state, these materials exhibit non-Fermi liquid properties and enter a mysterious pseudogap (PG) regime, characterized by the observation of multiple crossover PG temperatures T$^*$'s.

The exotic superconductors differ from ordinary Bardeen-Cooper-Schrieffer (BCS) superconductors in many other points. Thus, muon spin relaxation ($\mu$SR) measurements
of the magnetic field penetration depth revealed nearly linear relationship between T$_c$ and the superfluid density in high-T$_c$ cuprates and many other exotic superconductors that cannot be expected in BCS theory, but is typical for Bose-Einstein condensation (BEC) of preformed pairs\,\cite{Uemura}.
Bosonic scenario for high-T$_c$ cuprates\,\cite{RMP} has been elaborated by many authors, in particular, by Alexandrov (see, e.g., Ref.\,\cite{Alexandrov}) who considered real-space bipolarons. It is worth noting that many reasonable predictions of the  bipolaronic theory are valid for any local bosons irrespective of its microscopic mechanism.
 Numerous observations point to the possibility that high-T$_c$ cuprate superconductors may not be conventional BCS or BEC superconductors, but rather manifest a boson-fermion competition in a struggle for the electronic ground state, in particular, a competition between the two- and one-particle transport with resistivity $\propto$T and $\propto$T$^2$, respectively.

The most part of the current scenarios, including the Hubbard and  $t-J$-models, spin fluctuations, Alexandrov-Mott bipolarons   \cite{Alexandrov} consider cuprates to be homogeneous systems and ignore numerous   signatures of the electron and crystalline inhomogeneity\,\cite{Phillips}.   

Both the normal and high-transition-temperature (high-T$_c$) superconducting (SC) state in cuprates is believed to be electronically inhomogeneous, in particular,  due to a quenched disorder, arising from dopants and/or nonisovalent substitution. However, 
the dopant-induced impurity potential, seemingly being a natural source of electron inhomogeneity, varies widely among the cuprates that cannot explain observation of an universal, scaling  behavior evidencing for an intrinsic electronic tendency toward inhomogeneity in CuO$_2$ planes. This intrinsic propensity can be stimulated, firstly, by a local out-of-plane nonisovalent substitution, toward formation of in-plane universal inhomogeneity centers. Another stimulating factor of the intrinsic electronic inhomogeneity is related with a two-dimensionality and a competition and intertwinnig of charge, spin and orbital degrees of freedom  in CuO$_2$ planes. Concept of phase separation and percolation phenomena    \cite{PS,EmeryKivelson1993,EmeryKivelson1995nature,EmeryKivelson1995prl,Furrer}, stripes \cite{Tranquada,Bianconi,Zaanen},    large polarons \cite{Dionne,GBersuker}, nucleation of the mixed valence     PJT-(pseudo-Jahn-Teller) phase \cite{Moskvin-98} has appeared     to be very fruitful for explanation of many puzzling properties of cuprates. 
Furthermore, some authors\,\cite{Wiegman,Rodriguez,Skyrme} associate the anomalous properties of cuprates with quasi-2D structure of the active  $CuO_2$ layers and different  topological defects, or vortex-like solitons  to be  specific collective   excitation modes of the 2D vector fields.

The topological order inherent in the doped cuprate endows it with tremendous amount of robustness to various unavoidable "real-life" material complications \cite{Senthil}, such as impurities and other coexisting broken symmetries.
In general, such a complex, multiscale phase separation does challenge theories of high-temperature superconductivity that include complexity\,\cite{Bianconi2015}.


Recently\,\cite{Moskvin-11} we argued that an unique property of high-T$_c$ cuprates is related with a dual nature of the Mott insulating state of the parent compounds that manifests itself in two distinct energy scales for the charge transfer (CT) reaction: Cu$^{2+}$\,+\,Cu$^{2+}$\,$\rightarrow$\,Cu$^{1+}$\,+\,Cu$^{3+}$.
Indeed, the $d$\,-\,$d$ CT gap as derived from the optical measurements in parent cuprates such as La$_2$CuO$_4$ is 1.5-2.0\,eV while  the true (thermal) $d$\,-\,$d$ CT gap, or effective correlation  parameter $U_d$,  appears to be as small as 0.4-0.5\,eV. 
It means cuprates should be addressed to be \emph{d-d} CT unstable systems whose description implies accounting of the three many-electron valence states CuO$_4^{7-,6-,5-}$ (nominally Cu$^{1+,2+,3+}$) on an equal footing as a well-defined charge triplet. 
This allows us to introduce a minimal model for cuprates  with the on-site Hilbert space  reduced to only three states, three effective valence centers CuO$_4^{7-,6-,5-}$ (Cu$^{1+,2+,3+}$) where the electronic and lattice degrees of freedom get strongly locked together, and make use of the S=1 pseudospin formalism\,\cite{Moskvin-11,Moskvin-LTP,Moskvin-09,JPCM,Moskvin-SCES,Moskvin-JSNM-2016}. Such a formalism constitutes a powerful method to study complex phenomena in interacting quantum systems characterized
by the coexistence and competition of various ordered states\,\cite{Batista}. Overall, such a framework provides a simple and systematic methodology to predict and discover new kinds of orders. In particular, the pseudospin formalism provides the most effective way to describe different topological structures.

The paper is organized as follows. In Sec.\,1 we introduce a working model for the C$\mbox{u}$O$_4$ centers based on assumption that the three many-electron valence states CuO$_4^{7-,6-,5-}$ (nominally Cu$^{1+,2+,3+}$) form the “on-site” Hilbert space of the CuO$_4$ plaquettes. We restricted ourselves only by the consideration of the charge degree of freedom and have suggested simple geometrical vector representation for the on-site charge states. In Sec.\,2 we have  addressed an effectuve pseudospin Hamiltonian for the model cuprate. In Sec.\,3 we have  considered several simplified versions of the general Hamiltonian.   Sec.\,4 is devoted to description of unconventional localized topological structures typical for 2D S=1 (pseudo)spin systems. In Sec.\,5 we considered localized topological structures in a limiting case of the model, or so-called "negative-$U$"\, model. A brief summary is given in Sec.\,6.

\section{Working model of the C$\mbox{u}$O$_4$ centers}

Hereafter we consider the CuO$_4$ plaquette to be a main element of crystal and electron structure of high-T$_c$ cuprates and introduce a simplified toy  model with the “on-site” Hilbert space of the CuO$_4$ plaquettes  reduced to states formed by only three effective valence centers [CuO$_4$]$^{7-,6-,5-}$ (nominally Cu$^{1+,2+,3+}$, respectively). The centers are characterized by different conventional spin: s=1/2 for Cu$^{2+}$ center and s=0 for Cu$^{1+,3+}$ centers, and different orbital symmetry:$B_{1g}$ for the ground states  of the  Cu$^{2+}$ center,  $A_{1g}$   for the Cu$^{1+}$ centers, and the Zhang-Rice (ZR) $A_{1g}$ or more complicated low-lying non-Zhang-Rice  states for the Cu$^{3+}$ center.
Electrons of such configurations cannot be treated through a mean-field independent particle approach; therefore, their behavior is studied in terms of auxiliary neither Fermi nor Bose quasiparticles, representing combinations of atomic-like many-electron configurations\,\cite{Ashkenazi}.

The key problem that arises from the strong correlations in the normal state of the copper-oxide
superconductors is identifying the weakly interacting entities that make a particle interpretation of the current possible. All formulations of superconductivity are reduced to a pairing instability of such well-defined quasiparticles. However, there is good reason to believe that the construction of such entities may not be possible\,\cite{Phillips-2013}.
\begin{equation}
|\Psi\rangle = c_{-1}|Cu^{1+}\rangle + c_{0}|Cu^{2+}\rangle + c_{1}|Cu^{3+}\rangle 
,
	\label{function} 	
\end{equation}
Such an approach immediately implies introduction of the unconventional on-site  quantum superpositions 
that points to many novel effects related with local CuO$_4$ centers. 
Validity of such a model implies well isolated ground states of the three centers. This surely holds for the $^1A_{1g}$ singlet ground state of the Cu$^{1+}$ centers with nominally filled 3d shell whose excitation energy does usually exceed 2\,eV (see, e.g., Ref.\,\cite{Pisarev-2006} and references therein). 
The $b_{1g}\propto d_{x^2-y^2}$ character of the ground hole state in CuO$_4^{6-}$ cluster (Cu$^{2+}$ center) seems to be one of a few indisputable points in cuprate physics. A set of low-lying excited states with the energy $\geq$\,1.5\,eV includes bonding molecular orbitals with $a_{1g}\propto d_{z^2}$, $b_{2g}\propto d_{xy}$, and $e_{g}\propto d_{yz},d_{xz}$ symmetry, as well as purely oxygen nonbonding orbitals with $a_{2g}(\pi)$ and $e_u(\pi)$ symmetry (see, e.g., Refs.\,\cite{Moskvin-PRB-02,Moskvin-PRL}). 

In 1988 Zhang and Rice\,\cite{ZR} have proposed that the doped hole in a parent cuprate forms a Cu$^{3+}$ center with a well isolated local spin and orbital ${}^1A_{1g}$ singlet ground state which involves a phase coherent combination of the 2p$\sigma$ orbitals of the four nearest neighbor oxygens with the same $b_{1g}$ symmetry as for a bare Cu 3$d_{x^2-y^2}$ hole.  
The Zhang-Rice (ZR) singlet is a leading paradigm in  modern  theories of high-temperature superconductivity. However, both numerous experimental data and the cluster model calculations  suggest the involvement of some other physics  which introduces low-lying states into the excitation of the doped-hole state, or competition of conventional ZR singlet with another electron removal state(s), in particular, formed by the hole occupation of the oxygen nonbonding $a_{2g}(\pi)$ and $e_u(\pi)$ orbitals\,\cite{Moskvin-PRB-02,Moskvin-PRL,NQR-NMR,Moskvin-FNT-11,JETPLett-12}, the $a_{2g}(\pi)$ orbital to be the lowest in energy. 

Unified selfconsistent description of the charge, spin, and orbital degrees of freedom for CuO$_4$ centers with mixed valence is a hardly solvable task so we are forced to address simplified model approaches focusing on the quantum description of the charge degree of freedom that is responsible for superconductivity in cuprates. 

\subsection{The charge triplet model: S=1 pseudospin formalism}

To describe the diagonal and off-diagonal, or quantum local charge order we start with a simplified {\it charge triplet model} that implies a full neglect of spin and orbital degrees of freedom\,\cite{Moskvin-SCES}. Three charge states of the CuO$_4$ plaquette: a bare center $M^0$=CuO$_4^{6-}$, a hole center $M^{+}$=CuO$_4^{5-}$, and an electron center $M^{-}$=CuO$_4^{7-}$ are assigned to  three components of the S=1 pseudospin (isospin) triplet with the pseudospin projections  $M_S =0,+1,-1$, respectively.  Obviously, the model resembles that of so-called semi-hard-core bosons\,\cite{Moskvin-JETP-2015}, which are described by extended Bose-Hubbard model that assumes a truncation of the on-site Hilbert space to the three lowest occupation states n = 0, 1, 2 with further mapping to an anisotropic spin-1 model (see, e.g., Refs.\cite{Altman2002,Berg2008,Berg2010}). For 2D cuprates these states correspond to a "electron"\,  CuO$_4^{7-}$ (Cu$^{1+}$), "bare"\, CuO$_4^{6-}$ (Cu$^{2+}$), and  "hole"\, CuO$_4^{5-}$ (Cu$^{3+}$)  centers, respectively.

The S=1 (pseudo)spin algebra includes eight independent nontrivial pseudospin operators, three dipole and five quadrupole operators:
\begin{eqnarray}	
&&
	{\hat S}_z; 
	\;\;
	{\hat S}_{\pm}=\mp \frac{1}{\sqrt{2}}(S_{x}\pm iS_{y});
\\
&&
	{\hat S}_z^2;
	\;\;
	{\hat T}_{\pm}=\{S_z, S_{\pm}\};
	\;\;
	{\hat S}^2_{\pm}=\frac{1}{2}({\hat S}_x^2-{\hat S}_y^2 \pm i\{{\hat S}_x,{\hat S}_y\}) 
	.
\nonumber
\end{eqnarray}

One should note a principal difference between the s=1/2 and  S=1 quantum systems. The only on-site order parameter in the former case is an average spin moment $\langle S_{x,y,z}\rangle$, whereas in the latter one has five additional "spin-quadrupole", or spin-nematic order parameters described by traceless symmetric tensors
\begin{equation}
	Q_{ij}=\langle \frac{1}{2}\{S_{i},S_{j}\}-\frac{2}{3}\delta_{ij} \rangle .
\end{equation}
 Interestingly, that in a sense, the  $S=\frac{1}{2}$ quantum spin system is closer to a classic one  ($S\rightarrow \infty$) with all the order parameters defined by a simple on-site vectorial order parameter $\langle {\bf S}\rangle$  than the  S=1 quantum spin system with its eight independent on-site  order parameters.

It is worth noting that the three spin-linear (dipole) operators $\hat S_{x,y,z}$ and five independent
spin-quadrupole operators $Q_{ij}=\frac{1}{2}\{{\hat S_{i}},{\hat S_{j}}\}-\frac{1}{3} {\hat
{\bf S}}^{2}\delta _{ij}$ at S=1 form eight Gell-Mann operators being the
generators of the SU(3) group\,\cite{Nadya}.


To describe different types of  pseudospin ordering in a mixed-valence system
we have to introduce eight local (on-site) order parameters: two classical ($diagonal$) order parameters: $\langle S_z
\rangle$ being a "valence", or charge density with an electro-neutrality
constraint, and $\langle S_{z}^{2} \rangle$ being the density of polar centers $M^{\pm}$, or "ionicity", and six {\it off-diagonal}
order parameters. 
The {\it
off-diagonal} order parameters describe different types of the valence mixing.

It should be emphasized that for the S=1 (pseudo)spin algebra there are two operators: $S_{\pm}$ and $T_{\pm}=\{S_z, S_{\pm}\}$ that change the pseudo-spin projection by $\pm 1$, with slightly different properties
\begin{equation}
\langle 0 |\hat S_{\pm} | \mp 1 \rangle = \langle \pm 1 |\hat S_{\pm} | 0
\rangle =\mp 1, \label{S1}
\end{equation}
but
\begin{equation}
\langle 0 |\hat T_{\pm}| \mp 1 \rangle = -\langle \pm 1 |(\hat T_{\pm}| 0 \rangle =+1. \label{S2}
\end{equation}
It is worth noting the similar behavior of the both operators under the hermitian conjugation:
${\hat S}_{\pm}^{\dagger}=-{\hat S}_{\mp}$;  ${\hat T}_{\pm}^{\dagger}=-{\hat T}_{\mp}$.

The ${\hat S}_{\pm}^{2}$ 
operator changes the pseudospin projection by $\pm 2$ with the local order parameter 
\begin{eqnarray}
\langle S_{\pm}^{2} \rangle
&=&
\frac{1}{2}(\langle S_x^2-S_y^2\rangle \pm i\langle\{S_x,S_y\}\rangle )={}
\\
&&
{}= c_{+}^*c_{-} = c_x^2-c_y^2\pm 2i c_x c_y .
\nonumber
\end{eqnarray}
Obviously, this on-site off-diagonal order parameter is nonzero only when both $c_+$ and $c_-$ are nonzero, or for the on-site "electron-hole"\, $M^-$(Cu$^{1+}$)-$M^+$(Cu$^{3+}$) superpositions. It is worth noting that the ${\hat S}_{+}^{2}$ (${\hat S}_{-}^{2}$) operator
creates an on-site hole (electron) pair, or composite boson, with a kinematic constraint $({\hat S}_{\pm}^{2})^2$\,=\,0, that underlines its "hard-core"\, nature.


Both ${\hat S}_{+}$(${\hat S}_{-}$) and ${\hat T}_{+}$(${\hat T}_{-}$) can be associated with the single particle creation (annihilation) operators, however, these are not standard fermionic ones, as well as ${\hat S}_{+}^2$(${\hat S}_{-}^2$) operators are not standard bosonic ones. Nevertheless, namely $\langle S_{\pm}^{2} \rangle$ can be addressed as a local superconducting order parameter

The two operators, $S_{\pm}$ and $T_{\pm}$ are related with the two different types of a correlated single-particle transport, these  change the pseudospin projection by $\pm 1$. 
In lieu of these operators one may use  two novel operators: 
$
	{\hat P}_{\pm}=\frac{1}{2}({\hat S}_{\pm}+{\hat T}_{\pm});\,{\hat N}_{\pm}=\frac{1}{2}({\hat S}_{\pm}-{\hat T}_{\pm})\,, 
$
which do realize transformations Cu$^{2+}$$\leftrightarrow$Cu$^{3+}$ and Cu$^{1+}$$\leftrightarrow$Cu$^{2+}$, respectively. In other words, for parent cuprates these are the hole and electron creation operators, respectively.
The  boson-like  pseudospin raising/lowering operators
${\hat S}_{\pm}^{2}$ do
change the pseudo-spin projection by $\pm 2$ and define a local nematic order parameter 
\begin{equation}
	\langle S_{\pm}^{2} \rangle
	= \frac{1}{2}( \langle S_x^2-S_y^2\rangle \pm i \langle \{S_x,S_y\} \rangle ).
\end{equation}
This on-site off-diagonal order parameter with the $d$-type symmetry is nonzero only for the on-site $M^-$(Cu$^{1+}$)-$M^+$(Cu$^{3+}$) superpositions. It is worth noting that the ${\hat S}_{+}^{2}$ (${\hat S}_{-}^{2}$) operator
creates an on-site hole (electron) pair, or composite boson, with a kinematic constraint $({\hat S}_{\pm}^{2})^2$\,=\,0, that underlines its "hard-core"\, nature. 
 Obviously, the pseudospin nematic average $\langle S_{\pm}^{2} \rangle$ can be addressed to be a local complex superconducting order parameter: 
\begin{equation}
	\langle S_{\pm}^{2} \rangle = |\langle S_{\pm}^{2} \rangle| e^{\pm i\varphi} .
\end{equation}
Both ${\hat S}_{+}$(${\hat S}_{-}$) and ${\hat T}_{+}$(${\hat T}_{-}$) can be anyhow related with conventional single particle creation (annihilation) operators, however, these are not standard fermionic ones, as well as ${\hat S}_{+}^2$(${\hat S}_{-}^2$) operators are not standard bosonic ones.

It should be noted again that the pseudospin operators are not to be confused with real physical spin operators; they act in a  pseudo-space.

\subsection{Simple "geometrical"\, representation of the on-site charge states}

Making use of a simple classical representation of on-site spin states using arrows is a popular and useful method for describing spin structures through vector fields. However, such an approach works only for classical spins and, under certain limitations, for spin s=1/2. Indeed, for the classical spin all the on-site spin order parameters are derived through $\langle{\bf S}\rangle$, while for s=1/2 $\langle{\bf S}\rangle$ is the only local spin order parameter. 
At variance with s=1/2 systems for S=1 systems we have additional spin-quadrupole order parameters whose description cannot be realized within framework of a classical "single-arrow"\, representation. Nevertheless, hereafter  we propose a novel "geometrical"\, representation that allows us to selfconsistently describe all the on-site S=1 states and make use of the 2D vector fields to describe uniform and nonuniform configurations  for model 2D cuprate. In particular, the vector field patterns are of a great importance for physically clear representation of the complex topological structures.

Instead of the three $|1M\rangle $ states one may use the Cartesian basis set ${\bf \Psi}$, or $|x,y,z\rangle$:
\begin{equation}
	|10\rangle = |z\rangle\, , \;
	|1 \, {\pm} 1\rangle = \mp\frac{1}{\sqrt{2}}(|x\rangle \pm i|y\rangle ) ,
\end{equation}
so that the on-site wave function can be written in the matrix form as follows\,\cite{Nadya}:
\begin{equation}
	\psi=\pmatrix{c_1\cr c_2\cr
	c_3}=\pmatrix{R_1\exp(i\Phi_1)\cr R_2\exp(i\Phi_2)\cr
	R_3\exp(i\Phi_3)}\,;\qquad |\vec R|^2=1\, , \label{fun}
\end{equation}
 with ${\bf R}=\{\sin\Theta\cos\eta ,\sin\Theta\sin\eta ,\cos\Theta\}$. Obviously, the minimal number of dynamic variables describing an isolated on-site S=1 (pseudo)spin center
  equals to four, however, for a more general situation, when
  the (pseudo)spin system represents only the part of the bigger system, and we are
  forced to consider the coupling with the additional degrees of freedom,  one should consider all the five non-trivial parameters.

The pseudospin matrix  has   a very simple form within the $|x,y,z\rangle$ basis set:
\begin{eqnarray}
\langle i |\hat S_{k} | j \rangle =i \epsilon _{ikj}.
\end{eqnarray}

We start by introducing the following set of S=1 coherent states characterized by vectors $\bf a$ and $\bf b$ satisfying the normalization constraint\,\cite{Nadya} 
\begin{eqnarray}
|{\bf c}\rangle = |{\bf a},{\bf b}\rangle 
= {\bf c}\cdot{\bf \Psi} = ({\bf a} +i{\bf b})\cdot{\bf \Psi} ,
\label{ab}
\end{eqnarray}
where ${\bf a}$ and ${\bf b}$ are real vectors that are
arbitrarily oriented with respect to some fixed coordinate
system in the pseudospin space with orthonormal basis ${\bf e}_{1,2,3}$. 

The two vectors are related by the normalization condition, so the minimal number of
dynamic variables describing the S=1 (pseudo)spin system appears to be equal to four.
Hereafter, we would like to emphasize the $director$ nature of the ${\bf c}$
vector field: $|{\bf c}\rangle$ and $|-{\bf c}\rangle$ describe the physically
identical states.

It should be noted that in a real space the $|{\bf c}\rangle$ state corresponds to a quantum on-site superposition 
\begin{eqnarray}
|{\bf c}\rangle = c_{-1}|Cu^{1+}\rangle +c_0|Cu^{2+}\rangle +c_{+1}|Cu^{3+}\rangle \, .
\label{function1}
\end{eqnarray}
Existence of such unconventional on-site superpositions is a princial point of our model.

Below instead of $\bf a$ and $\bf b$ we will make use of a pair of unit vectors $\bf m$ and $\bf n$, defined as follows\,\cite{Knig}: 
\begin{equation}
	\bf a = \cos\varphi \, {\bf m} ,\; \bf b = \sin\varphi \, {\bf n} .
\end{equation}

For the averages of the principal pseudospin operators we obtain
\begin{eqnarray}
\langle {\bf S} \rangle 
&=& 
\sin2 \varphi \, [{\bf m} \times {\bf n}]  , 
\\
\langle \{S_{i},S_{j}\} \rangle 
&=& 
2 ( \delta_{ij} - \cos^2 \varphi \, m_{i} m_{j} - \sin^2 \varphi \, n_{i} n_{j} ).
\end{eqnarray}


\begin{figure*}[t]
\begin{center}
\includegraphics[width=14cm,angle=0]{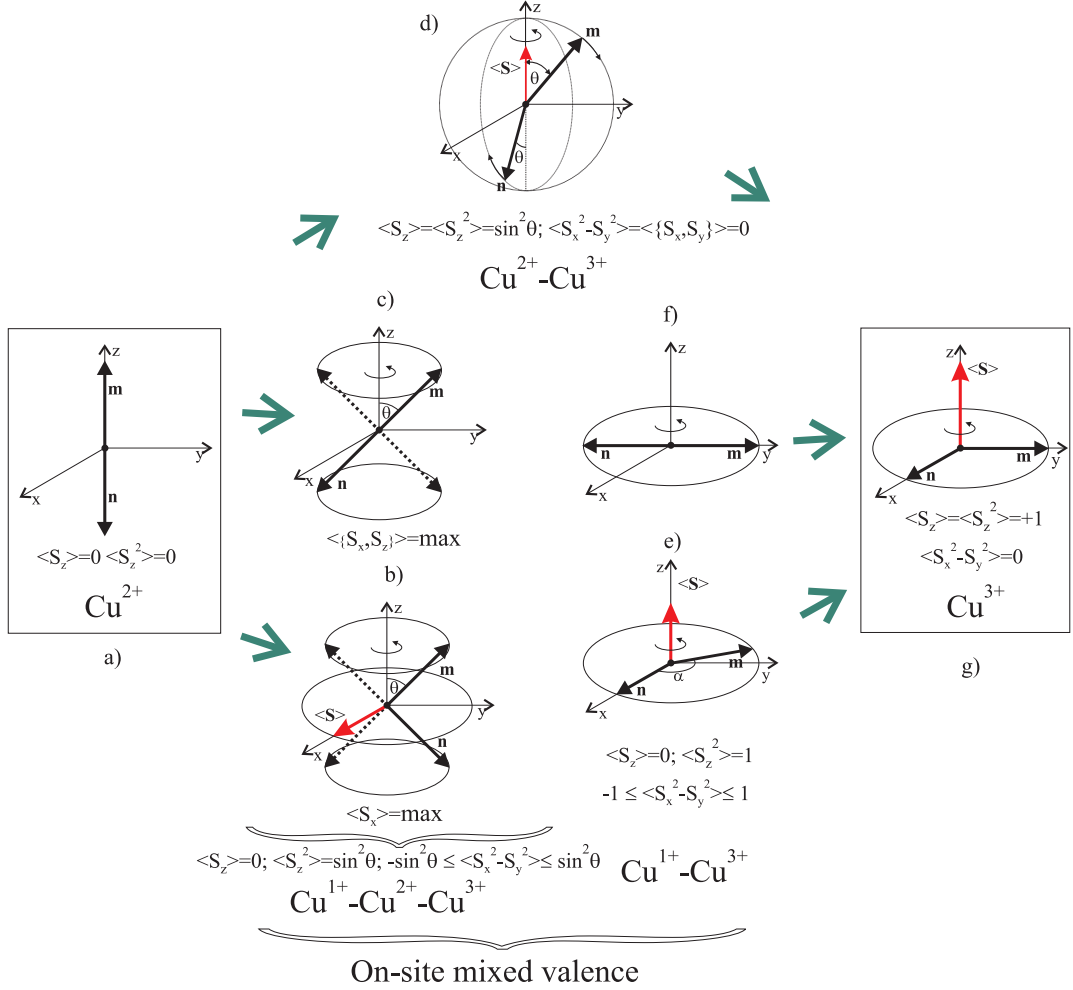}
\caption{(Color online) Cartoon showing orientations of the ${\bf m}$ and  ${\bf n}$ vectors which provide extremal values of different on-site pseudospin order parameters given $\varphi = \pi /4$ (see text for more detail).} \label{fig1}
\end{center}
\end{figure*}

Figure\,\ref{fig1} shows orientations of the ${\bf m}$ and  ${\bf n}$ vectors which provide extremal values of different on-site pseudospin order parameters given $\varphi = \pi /4$. The monovalent Cu$^{2+}$, or $M^0$ center, is described by a pair of $\bf m$ and $\bf n$ vectors directed along Z-axis with $|m_z|=|n_z|$\,=\,1. We arrive at the  Cu$^{2+}$-Cu$^{3+}$ ($M^0$-$M^+$) or Cu$^{2+}$-Cu$^{1+}$ ($M^0$-$M^-$) mixtures if turn $c_{-1}$ or $c_{+1}$, respectively, into zero. The mixtures  are described by a pair of $\bf m$ and $\bf n$ vectors whose projections on the XY-plane, ${\bf m}_{\perp}$ and ${\bf n}_{\perp}$, are of the same length and orthogonal to each other: ${\bf m}_{\perp}\cdot{\bf n}_{\perp}$\,=\,0, $m_{\perp}$\,=\,$n_{\perp}$ with $[{\bf m}_{\perp}\times{\bf n}_{\perp}]$\,=\,$\langle S_z\rangle$\,=\,$\pm \sin^2\theta$ for $M^0$-$M^{\pm}$ mixtures, respectively (see Fig.\,\ref{fig1}). 

It is worth noting that for "conical"\, configurations in Figs.\,1b-1d:
\begin{eqnarray}
&&
\langle S_z\rangle = 0 ,\; 
\langle S_z^2\rangle = \sin^2\theta ,\;
\langle S_{\pm}^2\rangle = - \frac{1}{2}\sin^2\theta \, e^{\pm 2i\varphi} ,
\nonumber
\\
&&
\langle S_{\pm}\rangle = -\frac{i}{\sqrt{2}}\sin2\theta \, e^{\pm i\varphi} ,\; 
\langle T_{\pm}\rangle = 0 ,
\end{eqnarray}
(Fig.\,1b)
\begin{eqnarray}
&&
\langle S_z\rangle = 0 ,\; 
\langle S_z^2\rangle = \sin^2\theta ,\;
\langle S_{\pm}^2\rangle = - \frac{1}{2}\sin^2\theta \,e^{\pm 2i\varphi} ,
\nonumber
\\
&&
\langle S_{\pm}\rangle=0 ,\;
\langle T_{\pm}\rangle=\mp \frac{1}{\sqrt{2}}\sin2\theta \,e^{\pm i\varphi} ,\; 
\end{eqnarray}
(Fig.\,1c)
\begin{eqnarray}
&&
\langle S_z\rangle = -\langle S_z^2\rangle = -\sin^2\theta ,\;
\langle S_{\pm}^2\rangle = 0 ,
\nonumber
\\
&&
\langle S_{\pm}\rangle = \langle T_{\pm}\rangle 
= \pm \frac{1}{2} e^{\mp i\frac{\pi}{4}} \sin2\theta \, e^{\pm i\varphi} ,
\end{eqnarray}
(Fig.\,1d). Figures 1e,f do show the orientation of ${\bf m}$ and ${\bf n}$ vectors for the local binary mixture Cu$^{1+}$-Cu$^{3+}$, and Fig.1g does for monovalent Cu$^{3+}$ center.
It is worth noting that for binary mixtures Cu$^{1+}$-Cu$^{2+}$ and Cu$^{3+}$-Cu$^{2+}$ we arrive at the same algebra of the ${\hat S}_{\pm}$ and ${\hat T}_{\pm}$ operators with $\langle S_{\pm}\rangle=\langle T_{\pm}\rangle$, while for ternary mixtures Cu$^{1+}$-Cu$^{2+}$-Cu$^{3+}$ these operators describe different excitations.
Interestingly that in all the cases the local Cu$^{2+}$ fraction can be written as follows:
\begin{equation}
	\rho (Cu^{2+})=1-\langle S_z^2\rangle = \cos^2\theta .
\end{equation}

\section{Effective S=1 pseudospin Hamiltonian}

Effective S=1 pseudospin Hamiltonian which does commute with the z-component of the total pseudospin  $\sum_{i}S_{iz}$ thus conserving the total charge of the system can be written to be a sum of potential and kinetic energies: 
\begin{equation}
{\hat H}={\hat H}_{pot}+{\hat H}_{kin}\, ,
\label{H}	
\end{equation}
where
\begin{equation}
	{\hat H}_{pot} =  \sum_{i}  (\Delta _{i}S_{iz}^2
	  - \mu S_{iz}) + \sum_{i<j} V_{ij}S_{iz}S_{jz} ,
\label{Hch}	   
\end{equation}
with a charge density constraint: 
\begin{equation}
\frac{1}{2N}\sum _{i} \langle S_{iz}\rangle =\Delta n	,
\label{Sz}
\end{equation}
where $\Delta n$ is the deviation from a half-filling ($N_{M^+}$\,=\,$N_{M^-}$). 
The first single-site term in ${\hat H}_{pot}$ describes the effects of a bare pseudo-spin splitting, or the local energy of $M^{0,\pm}$ centers and relates with the on-site density-density interactions, $\Delta = U/2$, $U$ being the correlation parameter. The second term   may be
related to a   pseudo-magnetic field $\parallel$\,$Z$ with $\mu$ being the hole chemical potential.  The third term in ${\hat H}_{pot}$ describes the inter-site density-density interactions.

In general the Hamiltonian (\ref{Hch}) describes so-called atomic limit of the model.

Kinetic energy ${\hat H}_{kin}={\hat H}_{kin}^{(1)}+{\hat H}_{kin}^{(2)}$ is a sum of one-particle and two-particle transfer contributions. 
In terms of ${\hat P}_{\pm}$ and  ${\hat N}_{\pm}$ operators the  Hamiltonian ${\hat H}_{kin}^{(1)}$ reads as follows: 
\begin{eqnarray} 
{\hat H}_{kin}^{(1)} 
&=& 
\frac{1}{2} \sum_{i\not= j} \big[ t^p_{ij} P_{i+} P_{j-} + t^n_{ij} N_{i+} N_{j-} +{}
\nonumber
\\
&&
{}+ \frac{1}{2} t^{pn}_{ij} ( P_{i+} N_{j-} + P_{i-} N_{j+} ) + h.c. \big]  .
\label{H1a}	
\end{eqnarray}
All the three terms here  suppose a clear physical interpretation. The first $PP$-type term describes one-particle hopping processes: 
$Cu^{3+}+Cu^{2+}\leftrightarrow Cu^{2+}+Cu^{3+}$,
that is  a rather conventional  motion of the hole $M^+$ ($Cu^{3+}$) centers in the lattice formed by $M^0$ ($Cu^{2+}$)-centers ($p$-type carriers, respectively) or the motion of the $M^0$ ($Cu^{2+}$)-centers in the lattice formed by hole $M^+$ ($Cu^{3+}$) centers ($n$-type  carriers, respectively). 
The second $NN$-type term describes 
one-particle hopping processes: $Cu^{1+}+Cu^{2+}\leftrightarrow Cu^{2+}+Cu^{1+}$,
that is  a rather conventional  motion of the electron $M^-$ ($Cu^{1+}$) centers in the lattice formed by $M^0$ ($Cu^{2+}$)-centers ($n$-type carriers) or the motion of the $M^0$ ($Cu^{2+}$)-centers in the lattice formed by electron $M^-$ ($Cu^{1+}$) centers ($p$-type  carriers).
These hopping processes are typical ones for heavily underdoped or heavily overdoped cuprates. 
The third $PN$ ($NP$) term in ${\hat H}_{kin}^{(1)}$ defines a very different one-particle hopping process:
$Cu^{2+}+Cu^{2+}\leftrightarrow Cu^{3+}+Cu^{1+}, Cu^{1+}+Cu^{3+}$,
that is the {\it local disproportionation/recombination}, or the {\it electron-hole pair creation/annihilation}. Interestingly, the term can be related with a local pairing as the $Cu^{1+}$ center can be addressed to be an electron pair (= composite electron boson) localized on the  $Cu^{3+}$ center or {\it vice versa} the $Cu^{3+}$ center can be addressed to be a hole pair (= composite hole boson) localized on the  $Cu^{1+}$ center. 
 
Hamiltonian ${\hat H}_{kin}^{(2)}$: 
\begin{equation}
  {\hat H}_{kin}^{(2)}=\sum_{i<j} t_{ij}^b(S_{i+}^{2}S_{j-}^{2}+S_{i-}^{2}S_{j+}^{2})\,,
  \label{H2}
\end{equation}
describes the two-particle (local composite boson) inter-site  hopping, that is the  motion of the electron (hole) center in the lattice formed by the hole (electron) centers, or the exchange reaction:
$Cu^{3+}+Cu^{1+} \leftrightarrow Cu^{1+}+Cu^{3+}$. In other words, $t^b_{ij}$ is the transfer integral for the local composite boson that governs the Bose condensation temperature. Its value is believed to be of a particular interest in cuprate physics

The magnitude of the effective  transfer
integral $t_{B}$  can be written as follows:
\begin{equation}
	t_{B}=  \langle 20| V_{ee}| 02\rangle - \sum_{11}\frac{\langle 20| \hat h| 11\rangle  \langle 11|  \hat h| 02\rangle }{\Delta_{dd}}\, ,
\end{equation}
where the first term describes a simultaneous tunnel transfer of the electron
pair due to Coulomb coupling $V_{ee}$ and may be called as a "potential"\,
contribution, while the second describes a two-step (20-11-02) electron-pair transfer
via   successive one-electron transfer due to the one-electron Hamiltonian, and may be called as "kinetic"\, contribution, $\Delta_{dd}$ is a Franck-Condon $d$-$d$ CT energy. As it is emphasized by P.W.
Anderson\,\cite{PWA} the value of the seemingly leading kinetic contribution to the composite boson
transport is closely related to the respective contribution to the exchange
integral, i.e. $t_{B}\approx$\,0.1\,eV in cuprates. From the other hand, this parameter determines so-called S-P separation for electron-hole dimers\,\cite{Moskvin-11} and can be reliably estimated from different optical data, both conventional optical conductivity\,\cite{MIR,Gruninger} and unconventional nonlinear optics\,\cite{Kishida,Ono-04,Maeda-04}. For different parent cuprates with corner-sharing CuO$_4$ plaquettes ( Sr$_2$CuO$_3$, Sr$_2$CuO$_2$Cl$_2$, YBa$_2$Cu$_3$O$_6$) the optical data point to surprisingly close values:  $t_{B}\approx$\,0.1\,eV.

These estimations point to  a promising  upper limit of $T\approx$\,1000\,K for the local bosonic superfluidity in cuprates and leaves a hope for the realization in the cuprates of the room-temperature superconductivity.

Similar to conventional spin systems the S\,=\,1 pseudospin formalism allows us to predict various types of  diagonal and off-diagonal long-range order and pseudospin excitations, including commensurate and incommensurate charge orders (pseudospin density waves), superfluid and supersolid phases, different topological excitations typical for 2D systems\,\cite{Moskvin-JETP-2015}.
However, at variance with typical spin systems, in particular, in 3d-oxides the pseudospin system appears to be strongly anisotropic one with an enhanced role of frustrative effects of the in-plane next-nearest neighbor couplings, inter-plane coupling, and different non-Heisenberg biquadratic interactions.
Despite the difference we can translate many results of the spin S\,=\,1 algebra to our pseudospin system. 
Turning to a classification of the possible homogeneous phases of the charge states of the model cuprates and its phase diagram we introduce $monovalent$ MV-1  (Cu$^{1+}$, Cu$^{2+}$, Cu$^{3+}$), $bivalent$ MV-2 (Cu$^{1+,2+}$, Cu$^{2+,3+}$, Cu$^{1+,3+}$),  and $trivalent$ MV-3 (Cu$^{1+,2+,3+}$)  phases in accordance with character of the on-site superpositions (\ref{function1}). 
Then, in accordance with the above nomenclature of spin phases  and  the charge triplet -- S\,=\,1 pseudospin correspondence we arrive at a  parent monovalent (Cu$^{2+}$) phase as an analogue of the quantum paramagnetic (QPM) phase to be a quantum analogue of the "easy-plane" phase,  the $XY^{13}$,  $XY^{123}$, $XY$-$Z_{FM}^{13}$, $XY$-$Z_{FM}^{23}$, $XY$-$Z_{FM}^{12}$,  $XY$-$Z_{FM}^{123}$,  $XY$-$Z_{AFM}^{13}$,  $XY$-$Z_{AFM}^{123}$, $XY$-$Z_{FIM}^{13}$,  $XY$-$Z_{FIM}^{123}$,  $Z_{FM}^{1}$, and $Z_{FM}^{3}$  phases as mono-, bi-, and trivalent analogues of respective spin phases. All the metallic phases with $XY^{13}$ and  $XY^{123}$ components do admit in principle the  pseudospin nematic  order $\langle S_{\pm}^2\rangle\not=$\,0  related with  the high-T$_c$ superconductivity (HTSC). In all the trivalent phases the superconducting order competes with a spin ordering. Moreover, in $XY$-$Z_{FIM}^{123}$ phase we deal with a competition of superconducting, spin, and charge orders. 
It is worth noting that the $XY$-$Z$ nomenclature does strictly reflect an interplay of kinetic  ($XY$-terms) and potential ($Z$-term)  energies, or itineracy and localization.

For undoped model cuprate with $\sum_i\langle S_{iz}\rangle$\,=\,0 (half filling) given rather large positive $\Delta >\Delta_1$ we arrive at insulating monovalent {\it quantum paramagnetic} $M^0$ (Cu$^{2+}$)-phase, a typical one for Mott-Hubbard insulators.  
In parent cuprates, such as La$_2$CuO$_4$, the Cu$^{2+}$ ions
form an antiferromagnetically (AF) coupled square lattice of
s = 1/2 spins, which could possibly realize the resonant
valence bond (RVB) liquid of singlet spin pairs. In the RVB
state the large energy gain of the singlet pair state, resonating
between the many spatial pairing configurations, drives strong
quantum fluctuations strong enough to suppress long range
AF order.
However, by lowering the $\Delta$ below $\Delta_1$ the undoped cuprate can be  turned first into metallic and superconducting $XY^{123}$ phase, and given $\Delta <\Delta_2$ into a fully disproportionated MV-2 system of electron $M^-$ and hole $M^+$ centers ($M^{\pm}$-phase) with 
$\langle S_{iz}^2\rangle$\,=\,1 (Fig.\,1), or {\it electron-hole Bose liquid} (EHBL)\,\cite{Moskvin-11,Moskvin-LTP,Moskvin-09,JPCM,Moskvin-98}.
There is no single particle transport: $\langle S_{\pm}\rangle =\langle T_{\pm}\rangle $\,=\,0,  while the bosonic one may exist, and, in common, $\langle S_{\pm}^2\rangle \not=$\,0.

Strictly speaking the S=1 pseudospin Hamiltonian (\ref{H}) describes an extended bosonic Hubbard model (EBHM) with truncation of the on-site Hilbert space to the three lowest occupation states n = 0, 1, 2, or the model of semi-hard-core bosons\,\cite{Moskvin-JETP-2015}. The EBHM Hamiltonian is a paradigmatic model for the highly topical field of ultracold gases in optical lattices, however, this is one of the working models to describe the insulator-metal transition and high-temperature superconductivity.
	
The pseudospin Hamiltonian (\ref{H}) can be generalized for cuprates to include spin and orbital degrees of freedom\,\cite{Moskvin-JSNM-2016}, in particular, to take into account an 	intra-plaquette charge nematicity.
Indeed, different orbital symmetry, $B_{1g}$ and  $A_{1g}$   of the ground states for  Cu$^{2+}$ and  Cu$^{1+,3+}$, respectively, unequivocally should result in a spontaneous orbital symmetry breaking accompanying the formation of the on-site superpositions with emergence of the on-site orbital order parameter of the $B_{1g}=B_{1g}\times A_{1g}$ ($\propto d_{x^2-y^2}$) symmetry. 
In frames of the CuO$_4$ cluster model the rhombic $B_{1g}$-type  symmetry breaking may be realized both by the $b_{1g}$-$a_{1g}$ ($d_{x^2-y^2}$-$d_{z^2}$) mixing for central Cu ion or  through the  oxygen subsystem either by emergence of different charge densities on the oxygens placed symmetrically relative to the central Cu ion and/or by the $B_{1g}$-type distortion of the CuO$_4$ plaquette resulting in different Cu-O separations for these oxygens.  The latter effect seems to be natural for Cu$^{1+}$ admixtures. Indeed, at variance with Cu$^{2+}$ and Cu$^{3+}$ ions the Cu$^{1+}$ ion due to a large intra-atomic $s$\,-\,$d$\,-\,hybridization does  prefer a dumbbell O-Cu-O linear configuration thus making large rhombic distortions of the CuO$_4$ cluster. Taking into account two energetically equivalent $B_{1g}$-type charge imbalance/distortions of the isolated CuO$_4$ plaquette in both cases we can introduce a dichotomic nematic variable that can be build in into effective pseudospin Hamiltonian.  
The STM\,\cite{nematic} and $^{17}$O NQR\,\cite{Haase} measurements of a static nematic order in cuprates support a charge imbalance between the density of holes at the oxygen sites  oriented along  $a$- and $b$-axes, however, there are clear signatures of the $B_{1g}$-type distortion (half-breathing mode) instabilities even in hole-doped superconducting cuprates which can be addressed to be a true "smoking gun"\,for electronic Cu$^{1+}$ centers. The two dynamically coexisting sets of CuO$_4$ clusters with different in-plane Cu-O interatomic distances have been really found by polarized Cu K-edge EXAFS in La$_{1.85}$Sr$_{0.15}$CuO$_4$\,\cite{Bianconi}. Giant phonon softening and line broadening of electronic origin of the longitudinal Cu-O bond stretching phonons near half-way to the zone boundary was observed in hole-doped cuprates (see, e.g., Ref.\,\cite{phonon} and references therein). Their amplitude follows the superconducting dome that supports our message about a specific role of electron-hole Cu$^{1+}$-Cu$^{3+}$ pairs in high-T$_c$ superconductivity.

Conventional spin s=1/2 degree of freedom can be build in our effective Hamiltonian, if we transform conventional Heisenberg spin exchange Cu$^{2+}$-Cu$^{2+}$ coupling as follows
\begin{equation}
	{\hat H}_{ex}=\sum_{i<j}{\hat I}_{ij}(\hat {\bf s}_i\cdot \hat {\bf s}_j)  ,
	\label{Hspin}
\end{equation}
where 
\begin{equation}
	{\hat I}_{ij}=(1-{\hat S}_{iz}^2)I_{ij}(1-{\hat S}_{jz}^2)
	\label{I}
\end{equation}
is an effective exchange integral, $(1-{\hat S}_{iz}^2)$ is a projection operator which picks out the s=1/2 Cu$^{2+}$ center, $I_{ij}$ is the conventional Cu$^{2+}$-Cu$^{2+}$ exchange integral. It is worth noting that $(1-\langle{\hat S}_{iz}^2\rangle )$ can be addressed to be a Cu$^{2+}$ spin density. 
Obviously, the spin exchange provides an energy gain to the parent antiferromagnetic insulating (AFMI) phase with $\langle {\hat S}_{iz}^2\rangle$\,=\,0, while local superconducting order parameter is maximal given $\langle {\hat S}_{iz}^2\rangle$\,=\,1. In other words, the superconductivity and magnetism are nonsymbiotic phenomena with competing order parameters giving rise to an intertwinning, glassiness, and other forms of electronic heterogeneities, especially considering the same order of magnitude for  $I_{ij}$ and $t^B_{ij}$. Simplified spin-pseudospin model  describing a spin-charge competition in atomic limit with an unconventional spin-charge phase separation in cuprates has been studied by us recently\,\cite{spin-charge}.

\section{Typical simplified S=1 spin  and pseudospin models}

Despite many simplifications, the
effective pseudospin Hamiltonian (\ref{H}) is rather complex, and represents one of the
most general forms of the anisotropic S=1 non-Heisenberg Hamiltonian.
Its real spin counterpart corresponds to an anisotropic S=1 magnet with a single-ion (on-site) and two-ion (inter-site bilinear and biquadratic) symmetric anisotropy in an external magnetic field under conservation of the total $S_z$.   
Spin Hamiltonian (\ref{H}) describes an interplay of the Zeeman,   single-ion and two-ion anisotropic terms giving rise to a competition of an (anti)ferromagnetic  order along Z-axis with an in-plane $XY$ magnetic order. Simplified versions of anisotropic S=1 Heisenberg Hamiltonian with bilinear exchange have been investigated rather extensively in recent years. Their analysis seems to provide an instructive introduction to description of our generalized pseudospin model.

\subsection{Anisotropic S\,=\,1 Heisenberg model with a single-ion anisotropy}

 Typical S\,=\,1 spin Hamiltonian with uniaxial single-site and exchange anisotropies reads as follows:

Effective pseudospin Hamiltonian (\ref{H}) can be reduced to 
a typical S\,=\,1 spin Hamiltonian with uniaxial single-ion and bilinear exchange anisotropies:
\begin{eqnarray}
	{\hat H} &=& \sum_{i<j} J_{ij} (S_{ix}S_{jx} + S_{iy}S_{jy} + \lambda S_{iz}S_{jz}) +{}
	\nonumber
	\\
	&&
	{}+ \sum_i D S_{iz}^2 - \sum_i h S_{iz}  ,
	\label{Hs}
\end{eqnarray}
if we neglect the two-particle transport term (\ref{H2}) and restrict ourselves by bilinear terms in the single-particle transport.
Correspondence with our pseudospin Hamiltonian points to $D=\Delta$, $J_{ij}=t_{ij}$, $\lambda J_{ij}=V_{ij}$.
Usually one  considers the antiferromagnet with $J>0$ since, in general, this is the case of
more interest. However, the Hamiltonian (\ref{Hs}) is invariant under the transformation $J,\lambda\rightarrow -J,-\lambda$ and a shift of the Brillouin zone ${\bf k}\rightarrow {\bf k}+(\pi ,\pi )$ for 2D square lattice. The system described by the Hamiltonian (\ref{Hs}) can be characterized by local (on-site) spin-linear order parameters $\langle {\bf S}\rangle$ and spin-quadratic (quadrupole spin-nematic) order parameters $Q^2_0=Q_{zz}=\langle S_z^2-\frac{2}{3}\rangle$ and $Q^2_{\pm 2}=\langle S_{\pm 1}^2\rangle$.

The model has been studied  rather extensively in recent years by several methods, e.g., molecular field approximation, spin-wave theories, exact numerical diagonalizations, nonlinear sigma model, quantum Monte Carlo, series expansions, variational methods, coupled cluster approach, self-consistent harmonic approximation,  and generalized SU(3) Schwinger boson representation (see, e.g., Refs.\,\cite{Sengupta2007,Hamer2010,Lapa2013} and references therein).

The spectrum of the spin Hamiltonian (\ref{Hs}) in the absence of external magnetic field changes drastically as $\Delta$
varies from very small to very large positive or negative values. A strong "easy-plane"\, anisotropy for large positive $\Delta >0$ favors a singlet phase where all the spins are in the $S_z = 0$   ground state. This quadrupole ($Q_{zz}$\,=\,-$\frac{2}{3}$) phase has no magnetic order, and is aptly referred to as a quantum paramagnetic phase (QPM), which is separated from the "ordered"\, state by a quantum critical point (QCP) at some $\Delta$\,=\,$\Delta_{1}$. 
A strong "easy-axis"\, anisotropy for large negative $\Delta \leq \Delta_2$, favors a spin ordering along $Z$, the "easy axis", with the on-site $S_z = \pm 1$ ($Z$-phase). The order parameter will be "Ising-like"\, and long-range (staggered) diagonal order will persist at finite temperature, up to a critical line T$_c(\Delta )$.   
For intermediate values $\Delta_1>\Delta >\Delta_2$ the Hamiltonian will have O(2) symmetry and the system is in a
gapless $XY$ phase. At $T=0$ the  O(2) symmetry will be spontaneously broken and the system will exhibit spin order in some direction.  Although there will be no ordered phase at finite temperature one expects a finite temperature Kosterlitz-Thouless transition. 
 At finite effective field $h_z$ but $\lambda$\,=\,1 the $XY$ phase transforms into a canted antiferromagnetic $XY$-$Z_{FM}$ phase, the spins acquire a uniform longitudinal component which increases with field and saturates at the fully polarized
(FP) state (all $S_z$\,=\,1, $Z_{FM}$ phase) above the saturation field $h_s$. 
However, at $D>$\,0  and $\lambda >$\,1 the phase diagram contains an extended spin supersolid or conical phase $XY$-$Z_{FIM}$ with ferrimagnetic $z$-order that does exist over a range of magnetic fields\,\cite{Sengupta2007,Hamer2010,Lapa2013}.

\subsection{"Negative"-$U$ model and its relevance for 2D cuprates}

At large negative values of the on-site correlation parameter $\Delta$\,=\,$U$/2 we arrive at the ground state of our model cuprate to be a system of electron CuO$_4^{7-}$ and hole CuO$_4^{5-}$ centers coupled by inter-site correlations and two-particle transport, while  single-particle transport described by ${\hat H}_{kin}^{(1)}$ is suppressed due to large value of the transfer energy. This electron-hole liquid is  equivalent to the lattice hard-core ($hc$) Bose system with an inter-site repulsion and can be termed as electron-hole Bose liquid (EHBL).  
Indeed, one may  address the electron $M^-$ center  to be a system of a local composite  boson ($e^2$) localized on the hole  $M^+$ center: $M^- = M^+ + e^2$. 
 For such a system, the pseudo-spin Hamiltonian (\ref{H}) can be mapped 
 onto the Hamiltonian of $hc$ Bose gas on a lattice (see Refs.\,\cite{RMP,bubble,bubble-2}  and references therein)
\begin{eqnarray}
	H_{hc}
	&=&
	-\sum\limits_{\langle ij \rangle} 
	t_{ij} {\hat P} ({\hat b}_{i}^{\dagger}{\hat b}_{j}+{\hat b}_{j}^{\dagger}{\hat b}_{i}) {\hat P} +{} 
 \nonumber
	\\ 
	&&
	{}+ \sum\limits_{\langle ij \rangle} V_{ij} n_{i} n_{j} - \mu \sum\limits_{i} n_{i},   
\label{hcB}
\end{eqnarray}
where ${\hat P}$ is the projection operator which removes double occupancy of
any site, ${\hat b}^{\dagger}({\hat b})$ are
the Pauli creation (annihilation) operators which are Bose-like commuting for
different sites $[{\hat b}_{i},{\hat b}_{j}^{\dagger}]=0,$ if $i\neq j,$  $[{\hat b}_{i},{\hat b}_{i}^{\dagger}]=1-2n_i$,
$n_i = {\hat b}_{i}^{\dagger}{\hat b}_{i}$; $N$ is a full number of sites, $\mu $  the chemical potential
determined from the condition of fixed full number of bosons $\sum_{i}\langle n_{i}\rangle $ or concentration $n=\frac{1}{N}\sum_{i}\langle n_{i}\rangle \in [0,1]$. The $t_{ij}$ denotes an effective transfer integral,  $V_{ij}$ is an
intersite interaction between the bosons. Hereafter, we'll consider only a nearest neighbor  boson-boson repulsion, $V_{ij}=V_{nn}=V>$\,0, and $t_{ij}=t_{nn}=t>$\,0. 
It is worth noting that near half-filling ($n\approx 1/2$) one might introduce the renormalization $n_i \rightarrow (n_i -1/2)$, or neutralizing background, that immediately provides the particle-hole symmetry. 
The model of hard-core bosons with an intersite repulsion is
equivalent to a system of s\,=\,1/2 spins   exposed to an external magnetic field
in the $z$-direction\,\cite{Matsuda-1970}. For the system with neutralizing background we arrive at an effective pseudo-spin Hamiltonian
\begin{equation}
H_{hc} = \sum_{\langle ij \rangle} J^{xy}_{ij} ({\hat s}_{i}^{+}{\hat s}_{j}^{-}+{\hat s}_{j}^{+}{\hat s}_{i}^{-}) +\sum\limits_{\langle ij \rangle}
J^{z}_{ij} {\hat s}_{i}^{z} {\hat s}_{j}^{z} - \mu \sum\limits_{i} {\hat s}_{i}^{z}, 
\label{spinB}
\end{equation}
where 
$J^{xy}_{ij}=2t_{ij}$, $J^{z}_{ij}=V_{ij}$, 
$\hat{s}_i^{-}= \frac{1}{\sqrt{2}}\hat{b}_i $, $\hat{s}_i^{+}=-\frac{1}{\sqrt{2}}\hat{b}_i^{\dagger}$, $\hat{s}_i^{z}=-\frac{1}{2}+\hat{b}_i^{\dagger}\hat{b}_i$,
$\hat{s}_i^{\pm}=\mp \frac{1}{\sqrt{2}}(\hat{s}_i^x \pm i\hat{s}_i^y)$.

Local on-site order is characterized by the three order parameters: $\langle {\hat s}^z\rangle$\,=$n$\,-\,$\frac{1}{2}$; $\langle {\hat s}^{\pm}\rangle$\,=\,$|\langle {\hat s}^{\pm}\rangle |e^{\pm i\varphi}$, related with the charge and superfluid degree of freedom, respectively.

 The EHBL model exhibits many fascinating quantum phases and phase
transitions. Early investigations\,\cite{RMP} point to the $T=0$ charge order (CO=$Z_{AFM}^{13}$), Bose
superfluid (BS=$XY$-$Z_{FM}^{13}$) and mixed (BS+CO=$XY$-$Z_{FIM}^{13}$) supersolid uniform phases with an Ising-type
melting transition (CO-NO=$Z_{AFM}^{13}$-$Z_{FM}^{13}$) and Kosterlitz-Thouless-type (BS-NO=$XY$-$Z_{FM}^{13}$-$Z_{FM}^{13}$) phase
transitions to a non-ordered normal fluid (NO=$Z_{FM}^{13}$) in 2D systems.
 At half-filling ($n_B=0.5, \Delta n=0$) given $t^b>V_{nn}$, $V_{nnn}$\,=\,0 the EHBL system obviously prefers a superconducting BS=$XY^{13}$ phase while at $t^b<V_{nn}$, $V_{nnn}$\,=\,0
it prefers an insulating checkerboard charge order CO=$Z_{AFM}^{13}$.

The mean-field phase diagram for the hard-core bosons is well-known (see, e.g., Ref.\,\cite{RMP}). First of all the MFA points to emergence of an uniform supersolid CO+SF phase with deviation away from half-filling. At a critical concentration: 
\begin{equation}
	\Delta n_c=\frac{1}{2}\left(\frac{V-2t}{V+2t}\right)^{\frac{1}{2}} 
\end{equation}
the supersolid phase does transform into  the SF phase at $T$\,=\,0.

\begin{figure}[t]
\includegraphics[width=8.5cm,angle=0]{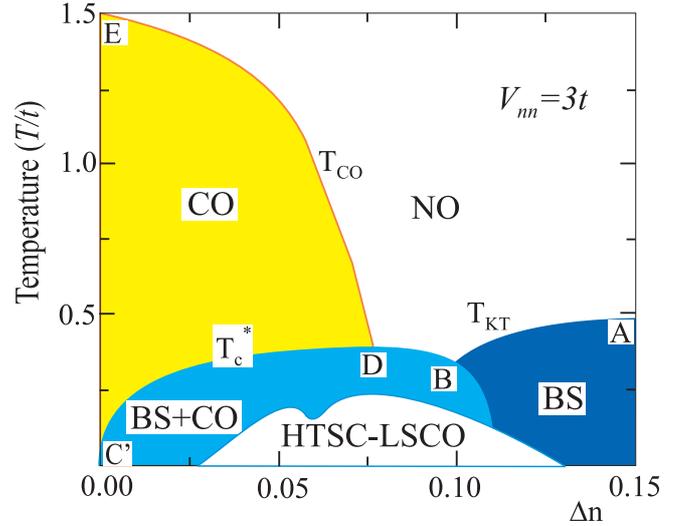}
\caption{Fig.\,2. (Color online) The QMC T-n phase diagram of the 2D hc-boson system (reconstruction of Fig.\,2 from Ref.\cite{Schmid}. Blank painting marks the HTSC phase for LSCO. } \label{fig2}
\end{figure}

 In Fig.\,\ref{fig2} we present the phase diagram of the square lattice hc-boson  model with the nearest neighbour ($nn$) transfer integral $t^b_{nn}=t$ (the Josephson coupling) and repulsion $V_{nn}=3t$, derived from the quantum Monte-Carlo (QMC) calculations by Schmid {\em et al.}\,\cite{Schmid}. Different filling points to CO phase, BS phase, and phase separated supersolid BS+CO phase. The AB line $T_{KT}(x)$ points to 2D Kosterlitz-Thouless phase transition; the C-B-D-C$^{\prime}$ line points to the first order phase transition; the D-E line $T_{CO}(\Delta n)$  which can be termed as the pseudogap onset temperature $T^*(\Delta n)$ points to the second order Ising kind melting phase transition CO-NO=$Z_{AFM}^{13}$-$Z_{FM}^{13}$ into a nonordered, or normal fluid phase.
It is worth noting that the QMC calculations\,\cite{Schmid} show that under doping away from half filling, the checkerboard solid undergoes phase separation: the superfluid (BS) and solid (CO) phases coexist but not as a single thermodynamic BS+CO phase. 

Simple uniform EHBL model truly reproduces many important aspects of the cuprate physics, in particular,  the pseudogap phenomenon as a result of the charge ordering and high values of the critical temperatures for superconducting transition. Indeed, the phase diagram in Fig.\,\ref{fig2} points to critical temperatures for 2D superconductivity as large as 0.5\,$t$ that is on the order of 500\,K, if we take into account reasonable estimations for the transfer integral in cuprates\,\cite{Moskvin-11}: $t\sim I\sim$\,1000\,K. Thus, the model is very promising for finding paths to room-temperature superconductivity\,\cite{Kresin}.  At the same time the model cannot explain a number of well-known properties, in particular, manifestation of the Cu$^{2+}$ valence states in doped cuprates  over wide doping range\,\cite{Johnston} and suppression of the superconductivity for overdoped cuprates. Such a behaviour cannot be derived from the EHBL scenario and  points to realization of the more complicated "boson-fermion"\, dual XY-Z$_{FIM}^{123}$ phase with coexisting spin and pseudospin (charge) orders in a wide doping range from parent to  overdoped compounds including all  the  superconducting phase.
The suppression of the superconductivity for the hole overdoped cuprates can be explained as a transition from the trivalent superconducting (M$^{123}$) phase to a bivalent nonsuperconducting M$^{23}$ phase. Indeed, the M$^{-}$=Cu$^{1+}$ centers could be energetically gainless under hole doping particularly for overscreened EH coupling.
Some properties of nonsuperconducting phases M$^{23}$ and M$^{12}$, or $XY$-$Z_{FM}^{23}$ and $XY$-$Z_{FM}^{12}$,  can be understood if we address limiting insulating phases $M^{+}$ or $M^{-}$ ($Z_{FM}^1$ or $Z_{FM}^3$) with precisely $M^{+}$ or $M^{-}$-centers on each of the lattice sites. In frames of the pseudospin formalism these phases correspond to fully polarized ferromagnetic states with $S_z^{tot}$\,=\,$\pm$\,N, where N is the number of Cu sites. Interestingly, in frames of the pseudospin formalism  the "heavily overdoped" \,$XY$-$Z_{FM}^{23}$ and $XY$-$Z_{FM}^{12}$ phases with $x\approx$\,1 can be represented as ferromagnets where the charge constraint is realized through the occurrence of $(1-x)N$ non-interacting pseudospin magnons ($\Delta S_z$\,=\,$\pm$\,1), that is Cu$^{2+}$ centers, obeying Fermi statistics due to s=1/2 conventional spin. These heavily overdoped cuprates could be addressed to be conventional Fermi liquids.  
Indeed, the standard  Fermi liquid theory hinges on the key assumption that although the electrons (holes) interact, the low-energy excitation spectrum stands in a one-to-one correspondence with that of a non-interacting system. In other words, the Fermi liquid behavior can be typical for 
overdoped $XY$-$Z_{FM}^{23}$ and $XY$-$Z_{FM}^{12}$ phases in a rather wide range of "overdoping".
The  phases are expected to manifest all the signatures of an $electron$ ($hole$) Fermi liquid 
with a large Fermi surface which contains $(1-x)$ electrons in the hole-doped cuprates such as La$_{2-x}$Sr$_x$CuO$_4$ or $(1-x)$ holes in the electron-doped cuprates such as Nd$_{2-x}$Ce$_x$CuO$_4$. 
However, this conventional Fermi liquid behavior can fail for hole or electron overdoped cuprates with a rather large content of Cu$^{2+}$ centers and the electron pairing due to disproportionation reaction Cu$^{2+}$+Cu$^{2+}$$\rightarrow$Cu$^{3+}$+Cu$^{1+}$  with formation of EH dimers and local bosons. In the framework of the pseudospin formalism we arrive at a binding of two pseudomagnons with $\Delta S_z$\,=\,-1 to a single bimagnon $\Delta S_z$\,=\,-2. In other words, we may think of the local boson as a long-lived virtual two-pseudomagnon bound state, or bimagnon, where the pseudomagnons are bound on the same site.
Electron pairing due to formation of the EH dimers, or CT exciton concept offers a fruitful insight into challenging issues of the copper oxide superconductors\,\cite{Gorkov,Gorkov-2,cofermion,Phillips}. The EH binding/unbinding energy $\Delta_{EH}$, that is the energy of the local boson binding in the EH dimer, can be identified with the pseudogap observed in the extended part of the cuprate phase diagram from heavily underdoped to overdoped systems. On the other hand, this energy  defines an effective gap for the thermal activation of hole carriers which has been found from the high-temperature Hall data in La$_{2-x}$Sr$_x$CuO$_4$\,\cite{Gorkov,Gorkov-2,Ando,Ono}.
Effective number of hole carriers derived from $R_H(T)$  were excelently fitted by a simple two-component formula where the first component is related with the temperature independent itinerant carriers, while the second one bears the activation character.  
Interestingly, the activation energy was shown to coincide with the ARPES measured excitation energies needed to transfer electron from the antinodal points (0,$\pi$), ($\pi$, 0) in the Brillouin zone   to the chemical potential\,\cite{Gorkov,Gorkov-2}.
 The equal energy for creation of an electron (ARPES) and a hole (thermal activation) can be understood only in terms of bound states for electron-hole pairs\,\cite{Gorkov,Gorkov-2}, EH-dimers, or condensation of the CT excitons whose coupling/decoupling energy is revealed in both types of experiments. These states, seen near antinodal points, according to ARPES, form quasi-periodic structures close to the double periodicity along the Cu-Cu bonds directions. 
The number of itinerant carriers rapidly increases with $x$\,\cite{Gorkov,Gorkov-2,Ando,Ono} that results in an effective screening of the  $\Delta_{EH}(x)$ parameter with a sharp fall of the "ionization"\, energy of the EH dimers from $\Delta_{EH}(x)\approx$\,0.5\,eV given $x$\,=\,0.01 up to values close to zero given $x>$\,0.2. Given zero values of the $\Delta_{EH}$ parameter the $XY$-$Z_{FIM}^{123}$ phase becomes energetically gainless as compared with $XY$-$Z_{FM}^{23}$ phase, hence we arrive at a QCP separating superconducting $XY$-$Z_{FIM}^{123}$ phase and nonsuperconducting $XY$-$Z_{FM}^{23}$ phase with a large Fermi surface (FS) and other attributes of a Fermi liquid. The onset of the pseudogap  below QCP naturally explains the  FS reconstruction with a number of unusual properties of the doped cuprates, such as the Fermi arc and/or pocket formation\,\cite{cofermion}.

It should be noted that the "negative-$U$"\, model is a limiting case of more complicated model with suppressed single-particle transport but with finite, positive or negative, values of the on-site correlation parameter $U$. Such a model was analyzed recently both within the mean-field approximation and quantum Monte-Carlo calculation\,\cite{Acta}.

\section{Topological defects in 2D S=1 pseudospin  systems}

\subsection{Short overview}
In the framework of our charge triplet  model the cuprates prove to be in the universality    class of the (pseudo)spin 2D systems whose description incorporates static or dynamic     topological defects to be natural element both of micro- and macroscopic physics. Like domain walls,  the vortices and skyrmions are stable for topological reasons. 
Depending on the structure of effective pseudo-spin Hamiltonian in 2D-systems the latter could correspond to either in-plane and out-of-plane vortices or skyrmions\,\cite{BP}. Under certain conditions either topological defects could determine the structure of the ground state. In particular, this could be a generic feature of electric multipolar systems with long-range multipolar interactions. Indeed, a Monte-Carlo simulation of a ferromagnetic Heisenberg model with dipolar interaction on a 2D square lattice $L\times L$ shows that, as $L$ is increased, the spin structure changes from a ferromagnetic one to a novel one with a vortex-like arrangement of spins even for rather small magnitude of dipolar anisotropy\,\cite{Sasaki}.

Quasi-classical continuous description of the 2D magnetic systems reveals their striking features, namely, the collective localized inhomogeneous states with nontrivial topology and finite excitation energy.
These include topological solitons\,\cite{BP,Voronov1983}, magnon drops\,\cite{Ivanov1977}, in- and out-of-plane vortex-antivortex pairs\,\cite{Gouva1989}, and various spiral solutions\,\cite{Borisov2001,Bostrem2002,Borisov2004}.
Basically these solutions have been obtained for the isotropic and anisotropic ferromagnet.


A vector representation is useful, if not a single instrument of a visual qualitative description of complex spin and pseudospin structures. A striking example of a single-vector representations are the N\'{e}el and Bloch domain walls in classic ferromagnets. Situation in quantum S=1 systems is more intricate, however, our "two-vector"\, (${\bf m}$, ${\bf n}$)-description of the on-site S=1 states can be successfully applied to predict and analyse different uniform and nonuniform, in particular, topological structures.
One of the most surprising is the prediction of the existence of unusual antiphase $180^{\circ}$-domain walls for parent cuprates.
Indeed, the bare on-site Cu$^{2+}$ state in parent cuprate is described by the collinear pair of vectors ${\bf m}$ and ${\bf n}$, or $-{\bf m}$ and $-{\bf n}$ directed along $Z$-axis. It means that two "parent"\, domains can be separated by a domain wall in which a collinear pair of vectors ${\bf m}$ and ${\bf n}$ can rotate by 180 degrees (see Fig.\,\ref{fig1}). Deviation from $Z$-axis 
corresponds to emergence of the on-site electron-hole Cu$^{1+}$-Cu$^{3+}$ component due to the gradual suppression of the bare parent Cu$^{2+}$ component until its complete disappearance in the domain wall center with the maximum value of the  electron-hole Cu$^{1+}$-Cu$^{3+}$ component. which ensures the maximum value of the electron-hole component and, accordingly, the maximum value of the modulus of the superconducting order parameter $\langle {\hat S}^2_{\pm}\rangle$. In other words, such a domain wall can be considered as a potential source of filamentary superconductivity. 

Interestingly, the domain wall structure is characterized by an uniform distribution of the mean on-site charge as $\langle {\hat S}_{z}\rangle $\,=\,0 for collinear (${\bf m}$, ${\bf n}$)-pair.   In other words, the domain wall structure and bare parent Cu$^{2+}$ monovalent (insulating) phase have absolutely the same distribution of the mean on-site charges. From the one hand, this point underlines an unconventional quantum nature of the on-site states indomain wall, while from the other hand it makes the domain wall textures to be invisible, in particular, for X-rays.

It is obvious that the formation of such domain walls in the parent cuprates is energetically unfavorable, however, the situation  changes radically with electron/hole doping due to the fact that the doping into a domain wall stabilizes the domain configuration\,\cite{bubble,bubble-2}. The antiphase domain wall in the parent cuprate appears to be a very efficient potential well for the localization of extra electron/hole pairs  
thus forming a novel type of a neutral or charged topological defect. We believe that the stripe structures in underdoped cuprates\,\cite{Tranquada,Bianconi,Zaanen} can somehow be associated with these antiphase domain walls.


Topological defects are stable non-uniform spin structures with broken translational symmetry and non-zero topological charge (chirality, vorticity, winding number). Vortices are stable states of anisotropic 2D Heisenberg Hamiltonian
\begin{equation}
	{\hat H}=\sum_{i<j}J_{ij}(S_{ix}S_{jx}+S_{iy}S_{jy}+\lambda S_{iz}S_{jz}) ,
	\label{H11}
\end{equation}
with the "easy-plane" anisotropy when the anisotropy parameter  $\lambda <1$.
Classical in-plane vortex ($S_{z}=0$) appears to be a stable solution of classical
Hamiltonian (\ref{H11}) at $\lambda  < \lambda _{c}$ ($\lambda _{c} \approx 0.7$
for square lattice). At  $1 > \lambda > \lambda _{c}$ stable solution corresponds
to the out-of-plane OP-vortex ($S_{z}\not= 0$), at which center the spin vector appears to be oriented along $z$-axis, and at infinity it arranges within $xy$-plane.  The in-plane vortex
is described by the formulas $\Phi=q\varphi$, $\cos\theta=0$. The $\theta(r)$
dependence  for the out-of-plane vortex cannot be found analytically. Both kinds of vortices have the energy logarithmically dependent on the size of the system.


The cylindrical domains, or bubble  like solitons with spins oriented  along the $z$-axis both at infinity and in the center (naturally, in opposite directions), exist for the "easy-axis" anisotropy $\lambda >1$. Their energy has a finite value.
 Skyrmions are general static solutions of classical continuous limit of the isotropic ($\lambda =1$) 2D Heisenberg ferromagnet, obtained by Belavin and Polyakov \cite{BP} from classical nonlinear sigma model. 
 Belavin-Polyakov skyrmion and out-of-plane vortex represent the simplest toy
model (pseudo)spin textures\,\cite{BP,Borisov}. 
 

The simplest skyrmion spin texture
looks like  a bubble domain in ferromagnet and consists of a vortex-like
arrangement of the in-plane components of spin with the $z$-component reversed
in the centre of the skyrmion and gradually increasing to match the homogeneous
background at infinity. 
 The spin distribution within such a classical skyrmion with a topological charge $q$ is given as follows\,\cite{BP}
\begin{equation}
\Phi=q\varphi +\varphi
_0;\quad\cos\Theta=\frac{r^{2q}-\lambda^{2q}}{r^{2q}+\lambda^{2q}}, \label{sk}
\end{equation}
where $r,  \varphi$ are polar coordinates on plane,  $q=\pm 1,\,\pm 2,...$ the chirality. For $q=1$, $\varphi_0$\,=\,0 we arrive at
$$
n_{x}=\frac{2r\lambda }{r^{2}+\lambda ^{2}}\cos\varphi ;\,
n_{y}=\frac{2r\lambda }{r^{2}+\lambda ^{2}}\sin\varphi ;\,
n_{z}=\frac{r^{2}-\lambda ^{2}}{r^{2}+\lambda ^{2}}\, .
$$
\begin{equation}
\label{sk1}
\end{equation}
In terms of the stereographic variables the skyrmion with radius  $\lambda $   and phase  $\varphi _{0}$  centered at a point $z_0$ is identified with spin distribution  $w(z)=\frac{\Lambda }{z-z_0}$, where $z=x+iy=re^{i \varphi }$  is a point in the complex plane, $\Lambda =\lambda e^{i\alpha}$.
For a multicenter skyrmion we have\,\cite{BP}
$$
w(z)=\cot\frac{\Theta}{2}\,e^{i\Phi}=\prod_i\left(\frac{z-z_j}{\Lambda}\right)^{m_j}\prod_j\left(\frac{\Lambda}{z-z_j}\right)^{n_j}\,,
$$
\begin{equation}
\end{equation}
where  $\sum m_i >\sum n_j$, $q=\sum m_j$.
Skyrmions are characterized by the magnitude and sign of its topological
charge, by its size (radius), and by the global orientation of the spin. The
scale invariance of skyrmionic solution reflects in that its energy $E_{sk}=4\pi |q|IS^2$ is proportional to  topological charge and does not depend on  radius and global phase\,\cite{BP}. Like domain
walls,  vortices and skyrmions are stable for topological reasons. Skyrmions cannot decay into other configurations because of this topological stability no matter
how close they are in energy to any other configuration.

In a continuous field model, such as, e.g., the nonlinear 
$\sigma$-model, the ground-state energy of the skyrmion does not
depend on its size\,\cite{BP}, however, for  the skyrmion  on
a lattice, the energy depends on its size. This must
lead to the collapse of the skyrmion, making
it unstable.
Strong anisotropic interactions, in particular, long 
range dipole-dipole interactions may,
in principle, dynamically stabilize the skyrmions in 2D
lattices\,\cite{Abanov,Ivanov2006,Galkina2009}.


Wave function of the spin system, which corresponds to a classical skyrmion,  is a product of spin coherent states \cite{Perelomov1986}. In case of spin  $S=\frac{1}{2}$
\begin{equation}
\Psi _{sk}( 0) =\prod\limits_{i}\lbrack \cos \frac{\theta _{i}}{2}e^{i\frac{\varphi _{i}}{2}}\mid \uparrow \rangle +\sin \frac{\theta _{i}}{2}e^{-i\frac{\varphi _{i}}{2}}\mid \downarrow \rangle \rbrack ,
\end{equation}
where $\theta _{i}=\arccos \frac{r_{i}^{2}-\lambda ^{2}}{r_{i}^{2}+\lambda ^{2}}$. Coherent state provides a maximal equivalence to classical state with minimal uncertainty of spin components.
The motion of such skyrmions has to be of highly quantum mechanical nature. However, this may involve a semi-classical percolation in the case of heavy non-localized skyrmions or variable range hopping in the case of highly localized skyrmions in a random potential.
Effective overlap and transfer integrals for quantum skyrmions are calculated analytically by Istomin and Moskvin \cite{Istomin}. The skyrmion motion has a cyclotronic character and resembles that of electron in a magnetic field.

The interest in skyrmions in ordered spin systems received much attention soon after the discovery of high-temperature superconductivity in copper oxides\,\cite{skyrmion-cuprates,skyrmion-cuprates-2,skyrmion-cuprates-3,skyrmion-cuprates-4,skyrmion-cuprates-5,skyrmion-cuprates-6,skyrmion-cuprates-7,skyrmion,skyrmion-2,skyrmion-3,skyrmion-4}. Initially, there was some hope that interaction of electrons and holes with spin skyrmions could play some role in superconductivity, but this was never successfully demonstrated. Some indirect evidence of skyrmions in the magnetoresistance of the litium doped lanthanum copper oxide has been recently reported\,\cite{skyrmion-LaLiCuO} but direct observation of skyrmions in 2D antiferromagnetic  lattices is still lacking.
In recent years the skyrmions and exotic skyrmion crystal (SkX) phases have been discussed in connection with a
wide range of condensed matter systems including quantum Hall effect, spinor Bose condensates and especially chiral magnets\,\cite{Bogdanov,Bogdanov-2,Nagaosa2013}.
It is worth noting that the skyrmion-like structures for hard-core 2D boson system were considered by Moskvin {\it et al.}\,\cite{bubble,bubble-2} in frames of the s=1/2 pseudospin formalism.


\subsection{Unconventional skyrmions in S=1 (pseudo)spin systems}

Different skyrmion-like topological defects for 2D (pseudo)spin S=1 systems as solutions of isotropic spin Hamiltonians were addressed in Ref.\,\cite{Knig}   and in more detail in Ref.\,\cite{Nadya}. 
In general, isotropic non-Heisenberg spin-Hamiltonian for the S=1 quantum
 (pseudo)spin systems should include both bilinear Heisenberg exchange term and
 biquadratic non-Heisenberg exchange term:
\begin{eqnarray}
	\hat{H}
	&=&
	-\tilde{J}_1 \sum_{i,\eta} \hat{\bf{S}}_i 	\hat{\bf{S}}_{i+\eta}
	-\tilde{J}_2\sum_{i,\eta} ( \hat{\bf{S}}_i \hat{\bf{S}}_{i+\eta} )^2=
	\label{ha1} 
	\\
	&=&
	-J_1\sum_{i,\eta} \hat{\bf{S}}_i \hat{\bf{S}}_{i+\eta}
	-J_2\sum_{i,\eta}\sum_{k\geq j}^3 (\{\hat{S}_k\hat{S}_j\}_i\{\hat{S}_k\hat{S}_j\}_{i+\eta}) ,
	\nonumber
\end{eqnarray}
where $J_i$ are the appropriate exchange integrals, $J_1=\tilde
J_1-\tilde J_2/2$, $J_2=\tilde J_2/2$, $i$ and $\eta$ denote the
summation over lattice sites and nearest neighbours, respectively.


Having  substituted  our trial wave function (\ref{ab}) to $\langle {\hat
H}\rangle$ provided $\langle \hat{\bf S}(1)\hat{\bf S}(2)\rangle =\langle
\hat{\bf S}(1)\rangle \langle \hat{\bf S}(2)\rangle $ we arrive at
 the Hamiltonian of the isotropic classical spin-1 model in the continual
  approximation as follows:
\begin{eqnarray}
	H
	&=&
	J_1\int d^2{\bf r}\left[\sum_{i=1}^{3}(\vec{\nabla}\langle S_i \rangle )^2 \right] +{}
	\nonumber
	\\
	&&
	{}+ J_2 \int d^2 {\bf r} 
	\left[ \sum_{i \leq	j=1}^{3} (\vec{\nabla} a_i a_j+\vec{\nabla} b_i b_j)^2 \right] +{}
	\nonumber
	\\
	&&
	{}+ \frac{4(J_2-J_1)}{c^2}\int |\langle \hat{\bf S}\rangle|^2d^2{\bf r} , 
	\label{ham7}
\end{eqnarray}
where $\langle \hat{\bf S}\rangle=2[\bf a\times\bf b]$. It should be noted that  the third
"gradientless" term in the Hamiltonian breaks the scaling invariance of the model. 

\subsubsection{Dipole (pseudo)spin skyrmions}

Dipole, or magnetic skyrmions as the solutions of bilinear Heisenberg
 (pseudo)spin Hamiltonian when $J_2=0$ were obtained in Ref.\,\cite{Knig} given the restriction
   $\bf a\perp\bf b$ and the lengths of these vectors were fixed.

The model  reduces to the nonlinear
 O(3)-model with the solutions for $\bf a$ and $\bf b$  described by the following formulas
 (in polar coordinates):
\begin{eqnarray}
	\sqrt{2}{\bf a}
	&=&
	({\bf e_z}\sin\theta - {\bf e}_{r}\cos\theta)\sin\varphi + {\bf e}_{\varphi}\cos\varphi ,
	\nonumber
	\\
	\sqrt{2}{\bf b}
	&=&
	({\bf e}_z\sin\theta - {\bf e}_{r}\cos\theta)\cos\varphi-
	{\bf e}_{\varphi}\sin\varphi .
	\label{knig}
\end{eqnarray}

For dipole "magneto-electric" skyrmions the $\bf m,\bf n$ vectors are assumed to be perpendicular to each other ($\bf m\perp \bf n$) and the (pseudo)spin structure is determined by the skyrmionic distribution (\ref{sk}) of the ${\bf l}=[{\bf m}\times{\bf n}]$ vector\,\cite{Knig}. In other words, the fixed-length spin vector $\langle {\bf S}\rangle =2[{\bf a}\times{\bf b}]$  is distributed in the same way as in the usual skyrmions (\ref{sk}).
However,
unlike the usual classic skyrmions, the dipole skyrmions in the S=1 theory 
have additional topological structure due to
 the existence of two vectors $\bf m$ and $\bf n$. Going around the center
 of the skyrmion the vectors can make $N$ turns around the
 ${\bf l}$ vector. Thus, we can introduce two topological
 quantum numbers: $N$ and $q$\,\cite{Knig}. In addition, it should be noted that
 $q$  number  may be half-integer.
 The dipole-quadrupole skyrmion is characterized by nonzero both pseudospin dipole order parameter $\langle \bf S\rangle$ with usual skyrmion texture (\ref{sk}) and quadrupole order parameters 
\begin{equation}
	\langle \{{\hat S}_{i}{\hat S}_{j}\}\rangle 
	= 2\langle {\hat S}_{i}\rangle \langle {\hat S}_{j}\rangle = l_i l_j .
\end{equation}

\subsubsection{Quadrupole (pseudo)spin skyrmions}

Hereafter we address another situation with purely biquadratic
  (pseudo)spin Hamiltonian ($J_1$=0) and treat the
    non-magnetic (``electric'') degrees of freedom.
    The topological classification of the purely electric
    solutions is simple because it is also based on the usage of subgroup
     instead of the full group. We address the solutions given
      $\vec a\parallel\vec b$ and the fixed
      lengths of the vectors, so we use for the
     classification the same subgroup as above.

After simple algebra the biquadratic part of the Hamiltonian can be reduced to the expression  familiar for nonlinear O(3)-model:
\begin{eqnarray}
	H_{bq}
	&=&
	J_2 \int d^2{\bf r} \left[ \sum_{i,j=1}^{3}(\vec{\nabla} n_i n_j)^2 \right] ={}
	\nonumber
	\\
	&=&
	2 J_2 |{\bf n}|^{2} \int d^2{\bf r} \left[ \sum_{i=1}^{3}(\vec{\nabla}n_i)^2 \right] ,
	\label{hm4} 
\end{eqnarray}
where
${\bf a}=\alpha {\bf n}, {\bf b}=\beta {\bf n}$, and $\alpha+i\beta= \exp (i\kappa)$,
$\kappa\in R$, $|{\bf n}|^2=$const. 
Its solutions are skyrmions, but instead of
the spin distribution in magnetic skyrmion we have  solutions with zero spin,
 but the non-zero distribution of five spin-quadrupole moments $Q_{ij}$, or
$\langle \{S_{i}S_{j}\}\rangle$ which in turn are determined
by the "skyrmionic"\, distribution of the ${\bf n}$ vector (\ref{sk})
with classical skyrmion energy: $E_{el}=16\pi qJ_2$. The distribution of the spin-quadrupole
moments $\langle \{S_{i}S_{j}\}\rangle$ can be easily obtained:
\begin{eqnarray}
&&
\langle S_z^2\rangle =\frac{4r^{2q}\lambda^{2q}}{(r^{2q}+\lambda^{2q})^2} ,\;
 \langle {\hat S}_{\pm}^2\rangle
=\frac{2r^{2q}\lambda^{2q}}{(r^{2q}+\lambda^{2q})^2}e^{\pm 2iq\varphi} ,
\nonumber
\\
&&
\langle {\hat T}_{\pm}\rangle =
-i\sqrt{2} \frac{(\lambda^{2q}-r^{2q})r^q\lambda^q}{(r^{2q}+\lambda^{2q})^2}e^{\mp i q\varphi} .
\label{qq}
\end{eqnarray}
One should be emphasized
that the distribution of five independent quadrupole order parameters for the
quadrupole  skyrmion are straightforwardly determined by a single vector field
${\bf m}({\bf r})$ (${\bf n}({\bf r})$) while $\langle \hat {\bf S}\rangle$\,=\,0.
\begin{figure}[t]
\begin{center}
\includegraphics[width=8.5cm,angle=0]{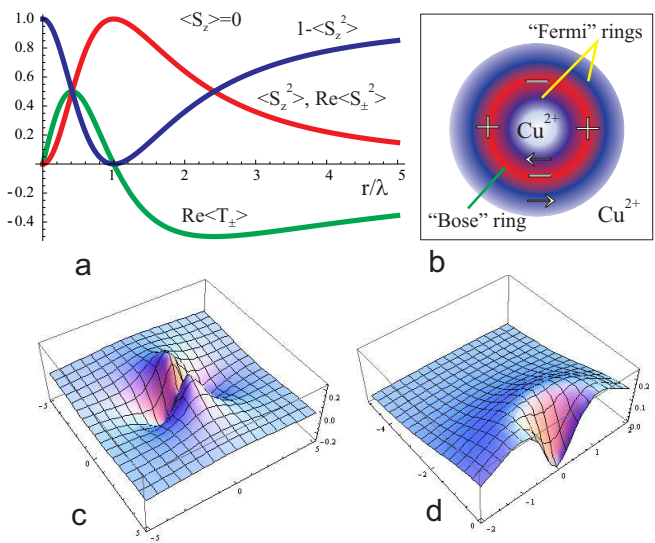}
\caption{(Color online) a) Radial distribution of the diagonal and off-diagonal nematic order parameters (the Cu$^{2+}$ spin density, the modulus of the SC order parameter, and the $T$-type order parameter) for a quadrupole pseudospin skyrmion (q=1): b) the ring shaped distribution of the Cu$^{2+}$ spin density and off-diagonal order parameters describing the Fermi- and Bose-like transport, arrows point to opposite sign of the $T$-type order parameter for the "internal"\, and "external"\, rings, the $\pm$ signs point to the signs of the $Re\langle {\hat S}_{\pm}^2\rangle $: c) and d) the spatial distribution of $Re\langle {\hat S}_{\pm}^2\rangle $ and  $\langle {\hat S}_{z}^2\rangle $, respectively. It should be noted the $d_{x^2-y^2}$ symmetry of the SC order parameter.}
\label{fig3}
\end{center}
\end{figure}

The quadrupole skyrmion supposedly can  be a typical topological charge excitation for parent or underdoped cuprates. 
Fig.\ref{fig3} demonstrates the radial distribution of different order parameters for the quadrupole skyrmion, the modulus of  the SC order parameter, the Cu$^{2+}$ spin density, and the $T$-type order parameter. We see a circular layered structure with clearly visible anticorrelation effects due to a pseudospin kinematics. Interestingly, at the center ($r=0$) and far from the center ($r\rightarrow\infty$) for such a skyrmion we deal with a parent Cu$^{2+}$ monovalent (insulating) state while for the domain wall center ($r=\lambda$) we arrive at a fully disproportionated  "superconducting" \, Cu$^{1+}$-Cu$^{3+}$ superposition whose weight diminishes with  moving away from the center. 
In other words, the ring shaped domain wall is an area with a circular distribution of the superconducting order parameter, or circular "bosonic"\, supercurrent. Nonzero $T$-type order parameter distribution points to a circular "fermionic" current with a puzzlingly opposite sign of the $\langle {\hat T}_{\pm}\rangle $ parameter for "internal"\, ($0<r<\lambda$) and "external"\, ($r>\lambda$) parts of the skyrmion.
Given the simplest winding number $q=1$ we arrive at the $d$-wave 
($d_{x^2 - y^2}$/$d_{xy}$ symmetry of the superconducting order parameter. 


First of all we should note that such a skyrmionic structure is characterized by an uniform distribution of the mean on-site charge as $\langle {\hat S}_{z}\rangle $\,=\,0, that is why it  can be termed as a neutral skyrmion.  Indeed, all over the skyrmion the ${\bf m}$ and ${\bf n}$ vectors form a collinear configuration, thus $\langle {\bf S}\rangle$ turns into zero.  In other words, the quadrupole skyrmionic structure and bare parent Cu$^{2+}$ monovalent (insulating) phase have absolutely the same distribution of the mean on-site charges. From the one hand, this point underlines an unconventional quantum nature of the quadrupole skyrmion under consideration, while from the other hand it makes the quadrupole skyrmion texture to be invisible, in particular, for X-rays. At the same time, the skyrmion has a well developed layered ring-shaped  distribution of the  $\langle {\hat T}_{\pm}\rangle $ and $\langle {\hat S}_{\pm}^2\rangle$ order parameters that points to its instability with regard to circular fermionic- and bosonic-like circular currents with maximal current density around the skyrmion radius. In this connection it is worth noting a scenario of the circulating charge currents for underdoped cuprates\,\cite{Varma,JETPLett-12}. 

\begin{figure}[t]
\begin{center}
\includegraphics[width=8.5cm,angle=0]{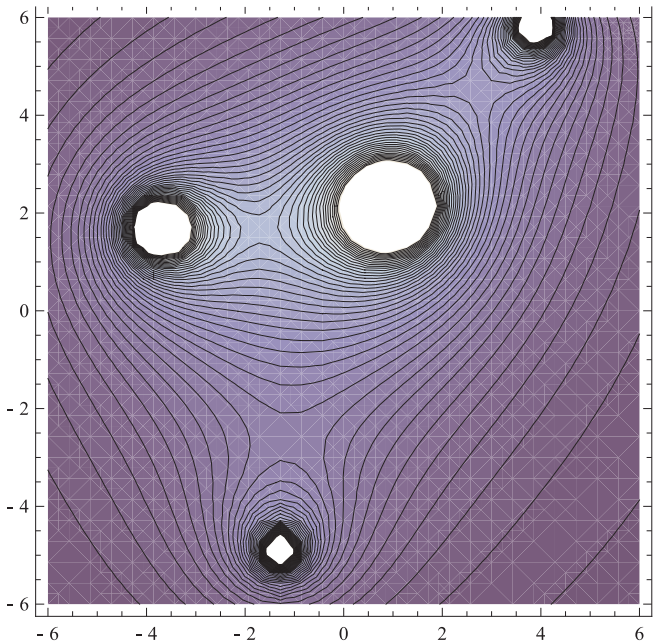}
\caption{ Contour plot of the in-plane distribution of the modulus of the off-diagonal SC order parameter $\langle{\hat S}^2_{\pm}\rangle$ for a multicenter quadrupole pseudospin skyrmion with a random distribution of the centers.}
\label{fig4}
\end{center}
\end{figure}

An interesting example of a topological inhomogeneity is provided by a multi-center skyrmion\,\cite{BP} which energy does not
depend on the position of the centers. The latter are believed to be addressed as  an additional degree of freedom, or positional order parameter. Fig.\,\ref{fig4} shows an example of the in-plane distribution of the modulus of the  SC order parameter $\langle{\hat S}^2_{\pm}\rangle$ for a multi-center quadrupole pseudospin skyrmion with a random distribution of the centers. An individual skyrmion in this multi-center entity can be characterized  by its position (i.e., the center of a skyrmionic texture), its size (i.e., the radius of domain wall), and the orientation of the in-plane components of pseudospin (U(1) degree of freedom).

The domain wall center of the quadrupole skyrmion provides maximal values of the pseudospin susceptibility $\chi_{zz}$, or charge susceptibility\,\cite{bubble,bubble-2}. It means the domain wall  appears to form a very efficient ring-shaped potential well for the charge carrier localization thus giving rise to a novel type of a charged topological defect. In the framework of the pseudospin formalism the skyrmion charging corresponds to a single-magnon $\Delta S_z$\,=\,$\pm$\,1 (single particle) or a two-magnon $\Delta S_z$\,=\,$\pm$\,2 (two-particle) excitations. It is worth noting that for large negative $\Delta$ the single-magnon (single-particle) excitations may not be the lowest energy excitations of the strongly anisotropic pseudospin system. Their energy may  surpass the energy of a two-magnon bound state (bimagnon), or two-particle, local boson-like, excitation, created at a particular site. Thus we arrive at a competition of the two types of charged quadrupole skyrmions. Such a charged topological defect can be addressed to be an extended  skyrmion-like mobile quasiparticle. However, at the same time it should'nt be forgotten that skyrmion corresponds to a collective state (excitation) of the whole system.

Skyrmionic scenario  allows us to make several important predictions for cuprates.
First, the  parent insulating antiferromagnetic monovalent Cu$^{2+}$ phase may be unstable with regard to  nucleation of a topological defect in the unconventional form of a single- or multi-center skyrmion-like object with  ring-shaped superfluid
regions.
The  parent Cu$^{2+}$ phase may gradually lose its stability under non-isovalent substitution (electron/hole doping), while a novel topological self-organized texture is believed to become stable.
 The most probable possibility is that every domain wall accumulates single
boson, or boson hole. Then, the number of centers in a multi-center skyrmion
nucleated with doping  has to be equal to the number of bosons/holes.   In such
a case, we anticipate the near-linear dependence of the total SC volume
fraction on the doping.
 Generally speaking, one may assume scenario
 when the nucleation of a  multi-center skyrmion  occurs spontaneously
 with no doping. In such a case we should anticipate the existence of neutral
 multi-center skyrmion-like object with equal number of positively and
negatively charged single skyrmions. However, in  practice, namely the
boson/hole doping is likely to be a physically clear driving force
for a nucleation of  a multi-center skyrmion-like self-organized
collective mode  which may be (not strictly correctly) referred to as
multi-skyrmion system akin in a quantum Hall ferromagnetic state of a
two-dimensional electron gas\,\cite{Green}.
It seems likely that for a light doping any  doped particle results in a nucleation
of a new single-skyrmion state, hence its density changes gradually with
particle doping.
Therefore, as long as the separation between skyrmionic centers is sufficiently
large so that the inter-skyrmion coupling is negligible, the energy of the
system per particle remains almost constant. This means that the chemical
potential remains unchanged with doping.
  Nucleation of the skyrmionic textures eventually leads to the destruction of the antiferromagnetic Ne\'{e}l ordering which is known to exist even at   very low doping. Furthermore, the skyrmion structure with insulating spin s=1/2 core isolated by spinless nonmagnetic Cu$^{1+}$-Cu$^{3+}$ ring-shaped domain wall from surrounding Cu$^{2+}$ entity provides a physically clear mechanism of the nucleation of a spin glass phase typical for underdoped cuprates. 
Furthermore, the nucleation of the unconventional quadrupole skyrmions does provide a physically clear mechanism for the unconventional vortex Nernst signal and local diamagnetism universally observed in many hole doped cuprates at the temperatures above T$_c$\,\cite{Nernst}.


Meanwhile we discuss the quadrupole skyrmion to be a classical solution of the continual isotropic model, however, this idealized object is believed to preserve their main features for strongly anisotropic (pseudo)spin lattice quantum systems. 
Both quantum effects and the discreteness of skyrmion texture can result
in  substantial deviations from the predictions of a classical model.
The continuous model is relevant for discrete lattices only if we
deal with long-wave length inhomogeneities when their size is much
bigger than the lattice spacing. In the discrete lattice the very
notion of topological excitation seems to be inconsistent. At the
same time, the discreteness of the lattice itself does not prohibit
from  considering the nanoscale (pseudo)spin textures whose
topology and spin arrangement is that of a skyrmion\,\cite{skyrmion-cuprates-6}.
It is worth to note that skyrmions cannot decay into other configurations because of the topological stability no matter
how close they are in energy to any other configuration.

The boson addition or removal in the half-filled ($n=1$) boson system  can be a driving force for a nucleation of  a  multi-center ``charged'' skyrmions.  Such   {\it topological} structures, rather than  uniform phases predicted by the mean-field approximation, are believed to describe the evolution of the EBHM systems away from half-filling. It is worth noting that the multi-center skyrmions one considers as systems of skyrmion-like quasiparticles forming skyrmion liquids and skyrmion lattices, or crystals (see, e.g., Refs.\,\cite{Timm,Green}).

\subsubsection{Dipole-quadrupole (pseudo)spin skyrmions}

In the continual limit for $J_1=J_2=J$ the Hamiltonian (\ref{ham7}) can be
transformed into the classical Hamiltonian of the fully $SU(3)$-symmetric
scale-invariant  model which can be rewritten as follows\,\cite{Nadya}:
\begin{eqnarray}
&&
H_{isotr} = 2J\int d^2{\bf r} 
\{(\vec{\nabla}\Theta)^2+\sin^2\Theta(\vec{\nabla}\eta)^2 +{}
\nonumber
\\
&&
{}+ \sin^2\Theta\cos^2\Theta
\left[ \cos^2\eta(\vec{\nabla}\Psi_1)^2+\sin^2\eta(\vec{\nabla}\Psi_2)^2 \right] +{}
\nonumber
\\[0.5em]
&&
{}+ \sin^4\Theta\cos^2\eta\sin^2\eta(\vec{\nabla}\Psi_1-\vec{\nabla}\Psi_2)^{2}\} ,
\label{iso}
\end{eqnarray}
where we have used the representation (\ref{fun}) and  introduced $\Psi_1=\Phi_1-\Phi_3,\Psi_2=\Phi_3-\Phi_2$. 
The topological solutions for the Hamiltonian (\ref{iso}) can be classified  at least by
three topological quantum numbers (winding numbers): phases $\eta,\Psi_{1,2}$
can change by $2\pi$ after the passing around the center of the defect.
The appropriate  modes may have very complicated topological structure due
to the possibility for  one defect to have several different centers
(while one of the phases $\eta,\Psi_{1,2,3}$ changes by $2\pi$ given
one turnover around one center $(r_1,\varphi_1)$, other phases may pass around other
 centers $(r_i,\varphi_i)$). It should be noted that for such a center the winding numbers
  may take  half-integer  values. Thus we arrive at a large variety of topological structures to be solutions of the model.  Below we will briefly address two simplest classes of such
 solutions.
 One type of skyrmions  can be obtained given the trivial
phases $\Psi_{1,2}$. If these are constant, the ${\bf R}$ vector distribution
(see ($\ref{fun}$)) represents the skyrmion described by the usual formula
(\ref{sk}).
  All but one topological quantum numbers are zero for this class of solutions.
  It includes both dipole and quadrupole solutions: depending
  on selected constant phases one can obtain both "electric" and different
  "magnetic" skyrmions. The substitution $\Phi_1=\Phi_2=\Phi_3$ leads to
  the electric skyrmion which was obtained above as a solution of more general
   SU(3)-anisotropic model. Another
   example can be $\Phi_1=\Phi_2=0,\Phi_3=\pi/2$. This substitution
   implies  ${\bf b}\Vert Oz,{\bf a}\Vert Oxy,{\bf S}\Vert Oxy$, and
   ${\bf S}=\sin\Theta\cos\Theta \{\sin\eta,-\cos\eta,0\}$.
   Nominally, this is the in-plane spin vortex with a varying length of
   the spin vector
\begin{equation}
	|S|=\frac{2r\lambda|r^2-\lambda^2|}{(r^2+\lambda^2)^2} ,
\end{equation}
which is zero at the circle $r=\lambda$, at the center $r=0$ and at
 the infinity $r\rightarrow \infty$,
and has maxima at  $r=\lambda(\sqrt 2 \pm 1)$. In addition to the non-zero
in-plane components of spin-dipole moment $\langle S_{x,y}\rangle$ this vortex is characterized by a non-zero
distribution of (pseudo)spin-quadrupole moments.
Here we would like to emphasize the difference between spin-1/2 systems in
 which there are such the solutions as in-plane vortices with the energy
  having a well-known logarithmic dependence on the size of the system
   and fixed spin length, and spin-1  systems in which the in-plane vortices
    also can exist but they may have a finite energy and a varying spin length.
     The distribution of quadrupole components associated with in-plane
      spin-1 vortex is non-trivial. Such solutions can be termed as
      "in-plane dipole-quadrupole skyrmions".

Other types of the simplest solutions with the phases
$\Psi_1=Q_1\varphi,\Psi_2=Q_2\varphi$ governed by two integer winding numbers
$Q_{1,2}$ and $\eta=\eta(r), \Theta=\Theta(r)$ are considered in Ref.\,\cite{Nadya}.

\section{Topological excitations in "negative-$U$"\, model}

\subsection{Quasi-classical approximation }

Let start with the Hamiltonian of the "negative-$U$"\, model in terms of the pseudospin $\vec{\sigma}$, $\sigma_z = \pm 1$:
\begin{equation}
		H =  
		- t \sum_{\left\langle ij\right\rangle} \big( \sigma_{i+} \sigma_{j-} + \sigma_{i-} \sigma_{j+} \big)
		+ V \sum_{\left\langle ij\right\rangle} \sigma_{iz} \sigma_{jz} 
	.
\end{equation}
Here $\sigma_\alpha$, $\alpha=x,y,z$ are Pauli matrices, $\sigma_{\pm}=(\sigma_x \pm i\sigma_y)/2$.
The $z$-component of the pseudospin describes the local density of composite bosons, 
so that antiferromagnetic $z$-$z$ exchange corresponds to the repulsive density-density interaction, 
while isotropic ferromagnetic planar exchange corresponds to the kinetic energy of the bosons. 
The constant total number of the bosons leads to the constraint on the total $z$-component of the pseudospin.
We define the $n$, as the density of the total doped charge counted from the state with a zero total z-component, 
or parent Cu$^{2+}$ state.
Then $n$ is the sum of $z$-components of the pseudospin:  $\sum_i \sigma_{iz} = nN$, 
where $N$ is the total number of sites.
If  $\rho$ is the density of \emph{hc} bosons, then  $n$ is the deviation from the half-filling:
$\rho = (1+n)/2$.

The energy functional 
$E = \left\langle \Psi \left| H \right| \Psi \right\rangle$ 
in a quasi-classical approximation with
\begin{equation}
	\left| \Psi \right\rangle
	= \prod_i
	\left(\cos \frac{\theta_i}{2} \, e^{-i\frac{\phi_i}{2}} \, | {+}1 \rangle
		+ \sin \frac{\theta_i}{2} \, e^{ i\frac{\phi_i}{2}} \, | {-}1 \rangle \right)	
	,
	\label{EnF}
\end{equation}
where $| \pm1 \rangle$ are the eigenfunctions of the $\sigma_{z}$ on $i$-th site, takes the form
\begin{eqnarray}
	\varepsilon 
	&=& 
	- \sum_{\langle ij \rangle} \sin\theta_i \, \sin\theta_j \, \cos(\phi_i-\phi_j)
	+{}
	\label{en}
	\\
	&&
	{}+ \lambda \sum_{\langle ij \rangle} \cos\theta_i \, \cos\theta_j
	- \xi  \Big(\sum_i \cos\theta_i   -   n N\Big) 
	.
	\nonumber
\end{eqnarray}
Here $\left\langle i,j\right\rangle$ denotes summation over nearest neighbors in a square lattice. 
The $\theta_i$ and  $\phi_i$ are the polar and azimuthal angles of the quasiclassical pseudospin vector at an $i$-th site.
We define
$\varepsilon = 2{E}/t$,
$\lambda = 2V/t$,
$\xi = 2\mu/t$,
where the chemical potential $\mu$ takes into account the bosons density constraint.

Hereafter, we introduce two sublattices $A$ and $B$ with the checkerboard ordering. 
The first sum in the Exp.(\ref{en}) has its lowest value if ${\cos(\phi_i-\phi_j)=1}$.
This allows us to make a simplifying assumption, that $\phi_A(\vec{r}) = \phi_B(\vec{r}) \equiv \phi(\vec{r})$. 
It is worth to note that this assumption is confirmed by the results of our numerical simulations.
We define functions 
$	u(\vec{r}) \equiv \theta_A(\vec{r}) $, 
$	v(\vec{r}) \equiv \theta_B(\vec{r}) $, 
and their combinations 
$f = \cos u \cos v$, $F = - f + \lambda f_{uv}$, 
where subscripts $u$ and $v$ denote derivatives with respect to these quantities.
Then the Euler equations in the continuous approximation take a compact form
\begin{equation}
	\left\{
	\begin{array}{l}
		f_{uv} \phi_{kk}  -  2 f_v  u_k  \phi_k  =  0
		,\\
		f_{uv} \phi_{kk}  -  2 f_u  v_k  \phi_k  =  0
		,\\
		\displaystyle
		F u_{kk}  +  F_u  u_k  u_k  -  f_u  \phi_k  \phi_k  -  4F_u  +  \xi \, \sin v  =  0
		,\\
		\displaystyle
		F v_{kk}  +  F_v  v_k  v_k  -  f_v  \phi_k  \phi_k  -  4F_v  +  \xi \, \sin u  =  0
		.
	\end{array}
	\right.
	\label{EqEulerSys1}
\end{equation}
Here we assume summation over pair indices.
These equations need to add the boson density constraint. 
With the relevant exchange constants, the equations (\ref{EqEulerSys1}) lead to the equations of Ref.\,\cite{Egorov2002}.


\subsection{The asymptotic behavior of localized solutions}
\label{sec-as}

The system (\ref{EqEulerSys1}) along with the boson density constraint has uniform solutions, 
$\phi=\phi_0$, $u=u_0$, $v=v_0$, 
that determine the well-known ground-state phase diagram\,\cite{RMP}  
of the \emph{hc} boson system in the mean-field approximation.

Given $\lambda<1$ or $n^2>(\lambda-1)/(\lambda+1)$ the ground state of the system is a superfluid (SF)
with
	$\cos u_0 = \cos v_0 = n$, 
	$\varepsilon_0 = -2 + 2 \left(\lambda + 1\right) n^2$.
Given $n^2<(\lambda-1)/(\lambda+1)$ the ground state is a supersolid (SS) with
	$\cos u_0 = n + z$,
	$\cos v_0 = n - z$,
where $z^2 = 1 + n^2 - 2|n|\lambda/\sqrt{\lambda^2 - 1}$, and
	$\varepsilon_0 = -2\lambda + 4|n|\sqrt{\lambda^2 - 1}$.
In all phases, the value of $\xi$ satisfies the regular expression $\xi_0 = \partial \varepsilon_0 / \partial n$.
When $\lambda>1$ and $n = 0$ the SS phase transforms into a conventional charge ordered (CO) phase with the checkerboard ordering.

\begin{figure}
	\centering
		\includegraphics[width=0.45\textwidth]{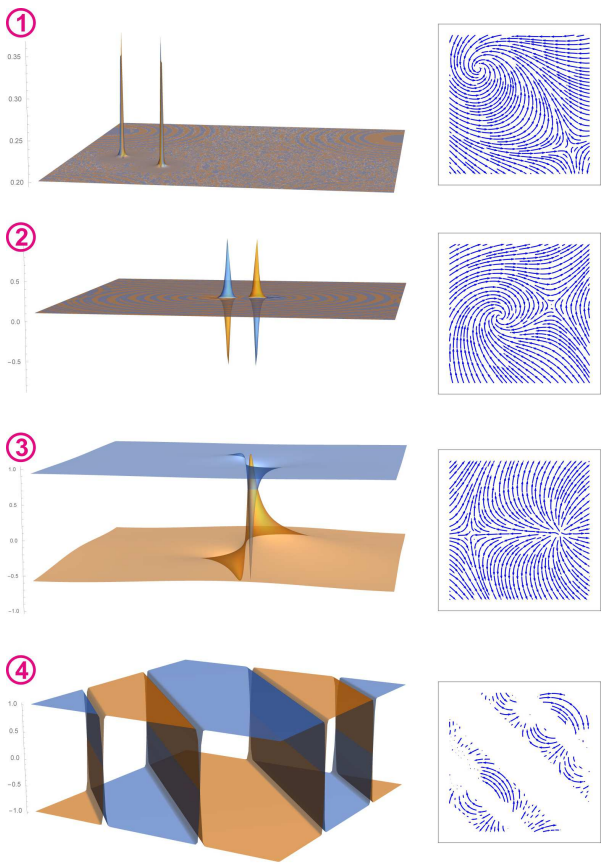}
	\caption{(color online)
	Nonuniform states in 2D system of charged hard-core bosons. 
	The left panels show 
	local charge density $n_i = \sigma_{zi} = \cos \theta_i$.
	The difference of sublattice states is clearly evident in the cases b, c and d.
	The right panels show 
	phase flow of planar components of the pseudospin, $\sigma_{xi}$ and $\sigma_{yi}$.
	The phase flow reveals the vortex-antivortex pair structure
	in the core of inhomogeneity of the local charge density in the cases a, b and c.
	The parameters of the model are: 
	a) $n=0.2$, $\lambda=0.5$ (SF phase);
	b) $n=0.1$, $\lambda=0.9$ (SF phase);
	c) $n=0.2$, $\lambda=1.5$ (SS phase);
	d) $n=0.0$, $\lambda=1.5$ (CO phase). 
	These sets of parameters are denoted with letters \emph{a}-\emph{d}  on the ground-state phase diagram in Fig.\ref{fig6}.
	} 
	\label{fig5}
\end{figure}

The equations (\ref{EqEulerSys1}) in the case of $\lambda <0$ have localized solutions with nonzero topological charge and finite energy \cite{BP,Voronov1983,Ivanov1977,Gouva1989,Borisov2001,Bostrem2002,Borisov2004}.
However, in our case $\lambda>0$, the numerical calculations with the conjugate gradient method for 
minimizing of the energy functional (\ref{EnF}) 
on the lattice 256$\times$256 indicate the existence of similar solutions, at least, as metastable states.
The results are shown in Fig.\ref{fig5}.
The actual stability of these solutions in our calculation was different. 
The SF-phase solutions, similar to that of in Fig.\ref{fig5}, cases \emph{a} and \emph{b}, quickly evolved to an uniform one. 
The SS- and CO-phase solutions, similar to that of in Fig.\ref{fig4}, cases \emph{c} and \emph{d}, 
retained its form for more than $10^6$ iterations.

We investigated the asymptotic behavior of localized solutions, suggesting that at $r\rightarrow\infty$ they have the form
$\phi(\vec{r}) = \phi_0 + \tilde{\phi}(\vec{r})$,
$u(\vec{r}) = u_0 + \tilde{u}(\vec{r})$, 
$v(\vec{r}) = v_0 + \tilde{v}(\vec{r})$, 
where $\tilde{\phi},\tilde{u},\tilde{v}\rightarrow0$ at $r\rightarrow\infty$. 
Hereinafter, the index 0 means the corresponding values for constant solutions. 
The linearized system (\ref{EqEulerSys1}) for the functions $\tilde{\phi},\tilde{u},\tilde{v}$ takes the form
\begin{equation}
	\left\{
	\begin{array}{l}
		\displaystyle
		f_{uv0} \,\tilde{\phi}_{kk}  =  0
		,\\
		\displaystyle
		F_0 \,\tilde{u}_{kk} 
		+ 4  F_0  \tilde{u} 
		+ 4  \left( - F_{uv0}  +  \frac{\xi_0}{4} \cos v_0 \right)  \tilde{v}  
		=  0
		,\\
		\displaystyle
		F_0 \,\tilde{v}_{kk}  
		+ 4  F_0  \tilde{v} 
		+ 4  \left( - F_{uv0}  +  \frac{\xi_0}{4} \cos u_0 \right)  \tilde{u}
		=  0
		.
	\end{array}
	\right.
	\label{EqEulerSys2}
\end{equation}

In the case of the SF and SS phases, the solutions for the first equation can be written as
\begin{equation}
	\tilde{\phi}(\vec{r}) 
	= \sum_{m=1}^{\infty}  \frac{c_m}{r^m}  \cos m(\varphi-\varphi_m)
	,
	\label{phimult}
\end{equation}
with $c_m$ and $\varphi_m$ determined by the boundary conditions. 
In the case of the CO phase, the first equation reduces to an identity since $f_{uv0}=0$.

In the case of the SF phase, the second and the third equations become independent Helmholtz equations 
for the ferro- and antiferro-type combinations 
$U = \tilde{u} + \tilde{v}$, $V = \tilde{u} - \tilde{v}$:
\begin{equation}
		U_{kk} + A_1 \, U  =  0
		,\quad
		A_1 = 4 \frac{(\lambda+1)(1-n^2)}{\lambda-(\lambda+1)n^2}
		;
		\label{linEqsSFU}
\end{equation}
\begin{equation}
		V_{kk} + A_2 \, V  =  0
		,\quad
		A_2 = 4 \frac{\lambda-1-(\lambda+1)n^2}{\lambda-(\lambda+1)n^2}
		.
		\label{linEqsSFV}
\end{equation}
The corresponding solutions have the form
\begin{eqnarray}
	\Phi(\vec{r},r_i)
	&=&
	\sum_{l=0}^{\infty}   a_{l}  K_l ( r/r_i )  \cos l ( \varphi - \alpha_{l} )
	,
	\label{Phi}
	\\
	\Psi(\vec{r},r_i)
	&=&
	\sum_{l=0}^{\infty} 
	\Big[ 
		b_{1l}  J_l ( r/r_i ) \cos l (\varphi-\beta_{1l})
		+{}
		\nonumber
	\\
	&&	
	{}+ b_{2l}  Y_l ( r/r_i ) \cos l (\varphi-\beta_{2l})
	\Big]
	,	
	\label{Psi}
\end{eqnarray}
where $K_l$ are the Macdonald functions, 
$J_l$ and $Y_l$ are the Bessel functions of the first and second kind, 
and $a_{l}$, $\alpha_{l}$, $b_{kl}$, $\beta_{kl}$, $k=1,2$ are some constants.
An analysis of the asymptotic behavior of solutions (\ref{Phi},\ref{Psi}) and the requirement that the omitted nonlinear terms in equations (\ref{EqEulerSys2}) are small as compared with the remaining linear terms point to $\Psi = 0$.
The account in the lowest order of the mixing with the function $\tilde{\phi}$ 
does not change $V(\vec{r})$ and gives additional term in $U(\vec{r})$
having asymptotic behavior:
\begin{equation}
	U_1(\vec{r}) \approx - \frac{n c_m^2}{2(\lambda+1)\sqrt{1-n^2}}  \frac{m^2}{r^{2m+2}}
	,
	\label{U1SF}
\end{equation}
where $m$ is the number that specifies first nonzero term in (\ref{phimult}).

Line $n^2 = \lambda/(\lambda+1)$ is the boundary of areas of the SF-phase with a different behavior of the  $U$ and $V$ functions
\begin{equation}
		n^2 > \frac{\lambda}{\lambda+1}: 
		\;
		U(\vec{r}) = \Phi(\vec{r},r_1) + U_1(\vec{r})
		,\;
		V(\vec{r}) = 0
		;
		\label{UVfin1}
\end{equation}
\begin{equation}
		n^2 < \frac{\lambda}{\lambda+1}:
		\;
		U(\vec{r}) = U_1(\vec{r})
		,\;\;
		V(\vec{r}) = \Phi(\vec{r},r_2)
		.
		\quad\;
		\label{UVfin2}
\end{equation}
Here we define characteristic lengths $r_i^{-2} = |A_i|$.

In the case of the SS phase, we need to define the ferri-type combinations: 
$\tilde{U} = A \tilde{u} + \tilde{v}$ and $\tilde{V} = A \tilde{u} - \tilde{v}$, 
where $A = -F_{uv0} + \frac{\xi_0}{4} \cos u_0$.
The equations (\ref{EqEulerSys2}) lead to Helmholtz equation for the $\tilde{U}$ function having solution
$\Psi(\vec{r},r_3)$,
$r_3^{-2} = 8$. 
As in previous case we have to put $\Psi = 0$.
The $\tilde{V}$ function obeys the Laplace equation. 
Taking into account the mixing  in the lowest order  with the function $\tilde{\phi}$ we come to the expressions as follows
\begin{equation}
		\tilde{U}(\vec{r})  
		= \frac{ c_m^2 }{8} \left( A f_{u0} + f_{v0} \right) \frac{m^2}{r^{2m+2}}
		,
\end{equation}
\begin{eqnarray}
		\tilde{V}(\vec{r}) 
		&=&
		\sum_{l=1}^{\infty}  \frac{C_l}{r^l}  \cos l (\varphi-\gamma_l)
		+{}
		\nonumber
		\\
		&&
		{}+
		\frac{ c_m^2 }{4} \left( A f_{u0} - f_{v0} \right) \frac{1}{r^{2m}}
		,
\end{eqnarray}
where $m$ is the number that specifies first non-zero term in (\ref{phimult}), and the expressions 
$f_{u0}=-\sin u_0 \cos v_0$, $f_{v0}=-\cos u_0 \sin v_0$ are determined by the expressions for the uniform solutions.

Similarly the case of the CO phase, 
we obtain
\begin{equation}
	\tilde{\phi}(\vec{r}) = 0
	,\quad
	U(\vec{r})= 0
	,\quad
	V(\vec{r}) = \Phi(\vec{r},r_4)
	,
	\label{COUV}
\end{equation}
where 	$r_4^{-2} = 4  (  \lambda - 1  )$.

The analysis of the asymptotic behavior of the localized states reveals qualitative differences of the finite energy excitations in the SF, SS, and CO phases. 

\begin{figure}
	\centering
		\includegraphics[width=0.45\textwidth]{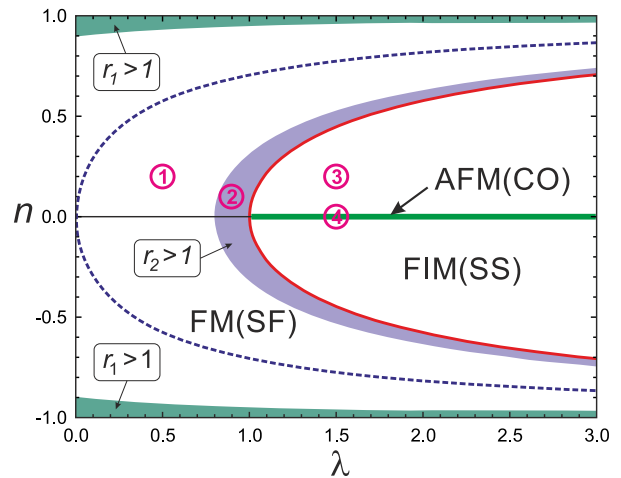}
		\caption{(color online)
		The ground state phase diagram of the \emph{hc} bosons in the mean-field approximation.
		The solid line corresponds to SF-SS phase boundary, $n^2 = (\lambda-1)/(\lambda+1)$. 
		The thick line at $\lambda>1$, $n=0$ shows CO phase.
		The dotted line, $n^2 = \lambda/(\lambda+1)$, 
		separates the two types of the asymptotic behavior in accordance with the expressions (\ref{UVfin1}) and (\ref{UVfin2}).
		In the shaded areas inside the SF phase region 
		the characteristic lengths satisfy to inequalities $r_i>1$. 
		The letters \emph{a}-\emph{d} in the circles correspond to the parameter sets in Fig.\ref{fig5}.
		}
	\label{fig6}
\end{figure}

In the SF phase an asymptotic of the polar angle of the pseudospin vector  is determined by the expressions (\ref{UVfin1}, \ref{UVfin2}).
When comparing these results with numerical calculations it is worth to note that the characteristic lengths obey to inequality $r_i<1$ in the most part of the phase diagram in the SF phase except for the areas indicated shadowed in Fig.\ref{fig6}, so the function $\Phi$ goes to zero value  
very fast 
with increasing of $r$. 
On the contrary, the asymptotic behavior of the azimuthal angle of the pseudospin (\ref{phimult}) has no characteristic scale. 
This means that in the SF phase the main excitations are almost in-plane vortex-antivortex pairs. 
They have well localized out-of-plane core of the ferro-type, with $\sigma_{zA}=\sigma_{zB}$, as shown in Fig.\ref{fig5}a,
and become the pure in-plane ones at $n=0$ in accordance with expression (\ref{U1SF}).
The same type of localized solutions was found by the authors of Ref.\,\cite{Gouva1989}.

For the \emph{hc} bosons, the polar angle is related with the density of bosons, while the azimuthal angle is responsible for the superfluid density, hence these states correspond to the excitation of the superfluid component with highly localized  heterogeneity of bosons density in the foci of the vortex-antivortex pairs.
In the shaded region in the SF phase near the border of the SF-SS phases in Fig.\ref{fig6}, the antiferro type vortices (see Fig.\ref{fig5}b), with $\sigma_{zA}\neq\sigma_{zB}$, begin to dominate, their inflation  is preceded by a change of the homogeneous ground state from SF to SS phase. 

The CO phase has no linear excitation of $\tilde{\phi}$. The characteristic lengths of the azimuthal excitations (\ref{COUV}) are small except the region near $\lambda=1$. 
This results in a high stability of the homogeneous CO phase. 
A typical picture of the nonuniform state shown in Fig.\ref{fig5}d is represented by linear domains of the CO phase. 
The non-zero values of the SF order parameter are realized within the domain walls, thus  
giving rise to appearance of a filamentary superfluidity for  \emph{hc} bosons.

In the SS phase the asymptotic behavior of the polar and azimuthal excitations is qualitatively the same without characteristic scales. 
Hence in this case there are skyrmion-like excitations as shown in Fig.\ref{fig5}c. 
For the \emph{hc} bosons these coherent states include both the excitations of the superfluid component and the boson density. 
In a center of skyrmion the difference $\sigma_{zA}-\sigma_{zB}$ has maximal magnitude, that corresponds to CO phase, 
and near there is a region where $\sigma_{zA}-\sigma_{zB}=0$, that corresponds to SF phase. 
So, the skyrmion-like excitations in the SS phase generate the topological phase separation. 
Note, that another type of instability in SS phase were also found by Quantum Monte Carlo calculations\,\cite{Batrouni2000}.

\subsection{Computer simulation of the domain structures}

Making use of a special algorithm for CUDA architecture for NVIDIA graphics cards that implies a nonlinear conjugate-gradient method to minimize energy functional and Monte-Carlo technique we have been able to directly observe formation of the ground state configuration for the 2D hard-core bosons  with lowering the temperature and its transformation with increase the temperature and boson concentration, allowing us to examine earlier  implications and uncover novel features of the phase transitions, in particular, look upon the nucleation of the odd domain structure, the localization of the bosons doped away from half-filling, and the phase separation regime. 
The accuracy of numerical calculations has been limited making it possible to reproduce the effect of  minor inhomogeneities common to any real crystal.

We started with the hard-core boson Hamiltonian (\ref{hcB}) on a 256$\times$256 square lattice at half-filling ($n\,{=}\,1/2$)  
given  the value of the inter-site repulsion $V_{nn}=V=3t_{nn}$, which is the typical one for many papers with QMC calculations for 2D hard-core bosons\,\cite{Batrouni2000,Hebert2001,Schmid}. It should be noted that hereafter we follow the notations in these papers.

First, we addressed the formation of the phase state under the annealing (thermalization) procedure. The algorithm starts initially with $T$ set to a high value $T\sim 2T_{cr}$. The annealing is accompanied by formation a fragile unstable CO domain structure with antiphase 180$^{\circ}$  domain walls whose center is characterized by a large nonzero SF order parameter which is suppressed  as one runs deep into the CO domain thus suggesting the presence of a fragile filamentary superfluidity (FLSF) nucleated at the antiphase domain walls. 
The term "filamentary superfluidity" is a full analogue of a more familiar term "filamentary superconductivity". Here, in the paper, "filamentary superfluidity" is related with an antiphase domain wall (the wall between two CO domains) which is characterized by a nonzero superfluid order parameter.

 \begin{figure*}[t]
\begin{center}
\includegraphics[width=16cm,angle=0]{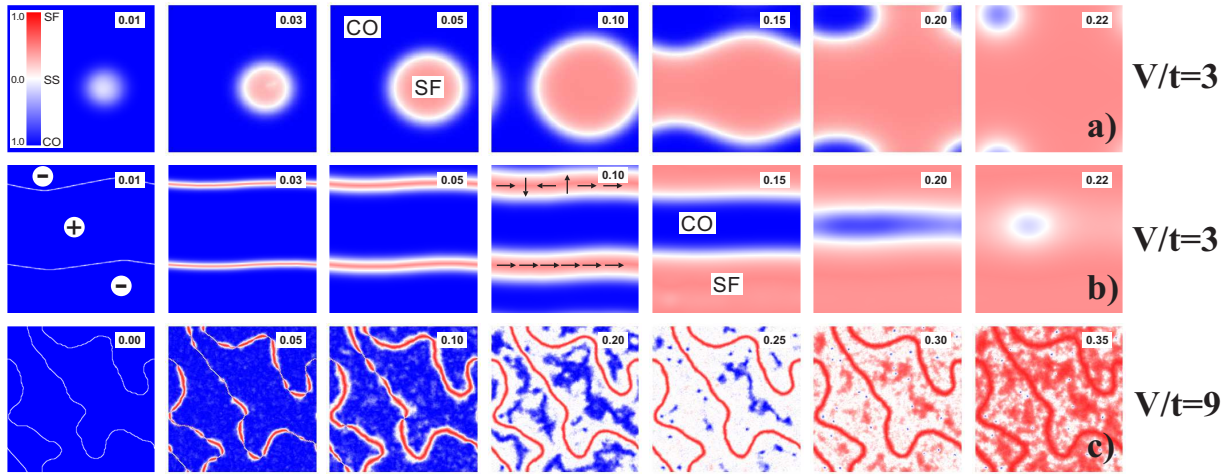}
\caption{(Color online) Evolution of the hc-boson ground state configuration under doping away from half-filling. a) At $\Delta n$\,=\,0.01 one observes a sudden nucleation of a rather large "blob"  composed of a SF core and a ring shaped SS boundary that does accommodate all the "injected" bosons. Under further deviation from half-filling the blob grows up to a full CO-SF phase transformation close to a critical value of $\Delta n$\,=\,0.225; b) Doped bosons do localize in narrow FLSF DWs  of the striped CO phase leading to their broadening. In a well developed phase separation regime we arrive at a system of nearly parallel CO and SF domains separated by the SS DWs. The "plus"\, and "minus"\, signs point to different CO domains. Orientation of the phase angle $\varphi$ within domain walls is demonstrated schematically for $\Delta n$\,=\,0.1.; c)  At $\Delta n$\,$<$\,0.01 the doped bosons do localize in the center of the narrow DWs breaking the FLSF without any visible transformation of the domains. The regular DW shape breaks under further doping, these are nonuniformly "swelled" with the emergence and rise of widenings. Step by step these widenings and "blobs" nucleated inside domains spread until these cover all lattice. The CO domain topology survives up to very high doping. Different color in a), b), c) does highlight the value of the order parameters. } 
\label{fig7}
\end{center}
\end{figure*}

 Typically for small and moderate anisotropy the annealing is finished by formation a system of domains with closed-loop domain walls which quickly collapse thus setting an uniform single-domain CO ground state with a hardly noticeable remnant inhomogeneity. At the lowest temperatures we can form an almost ideal charge ordered checkerboard structure at half-filling that does not modify with increasing the temperature up to $T_{CO}$. Weak deviation away from half-filling in such a case  gives rise to nucleation of two types of topological defects. Unconventional small nanoscopic defects with an effective radius of several lattice separations accommodate a single boson and are characterized by a strong distortion of the CO order extended on the three-four coordination spheres with emergence of the local superfluid order. Large topological defects (droplets, "blobs") having mainly a circular shape can accommodate many bosons. These   are comprised of a superfluid core and a ring shaped supersolid boundary. Computer simulation reveals occurrence of a critical radius for stability of the "large" cylindrical defects. Given the increasing doping we arrive at the growing volume fraction of the large defects, the change of their shape and their confluence up to a full phase transformation. At the same time, it is worth noting the persistence of the decreasing volume fraction of the checkerboard charge order up to very large doping.

However, systematic studies have indicated that in  some cases there occurs a low-temperature CO domain structure with stable stripe-like disconnected (within our lattice size)  domain walls oriented along main lattice axes. Along with a simple uniform ("ferromagnetic") SF phase parameter distribution these 1D walls can have unconventional multidomain topological structure of the SF phase order parameter with a high density of 2$\pi$ domain walls separating the 1D phase domains.

Evolution of the "uniform"\, and "striped"\, hc-boson ground state configurations under doping away from half-filling is shown in Fig.\,\ref{fig7}a and Fig.\,\ref{fig7}b, respectively, for a "moderate"\, anisotropy  $V=3t$.

Minor deviation away from half-filling at $\Delta n\leq$\,0.01 practically does not lead to visible effects by slightly perturbing the remnant inhomogeneity of the initial "uniform"\, state. However, at $\Delta n\approx$\,0.01 we observe a sudden nucleation of rather large topological defect(s) (droplets, "blobs") having mainly a circular shape that  can accommodate all the "injected"\, bosons thus making the surrounding CO phase more uniform. These droplets   are comprised of a superfluid core and a ring shaped supersolid boundary. Given the increasing doping we arrive at a well developed phase separation with the growing volume fraction of the large defects, the change of their shape and their confluence up to a full CO-SF phase transformation close to a critical value of $\Delta n_{cr}\approx$\,0.22.

Evolution of the "striped"\, CO phase with a filamentary superfluidity under deviation away from half-filling goes in different scenario  because of the doped bosons do localize in the center of the narrow domain walls leading to their uniform broadening up to formation of SF domains. Interestingly, the structure of the final SF phase in this case depends on the initial topological structure of the SF phase parameter ($\varphi$) within 1D domain walls. In Fig.\,\ref{fig7}b we started with the two 1D domain walls with an uniform distribution of the SF phase order parameter for lower wall and with 2$\pi$ domain wall separating the 1D phase domains for the upper wall. Orientation of the phase angle $\varphi$ within domain walls is demonstrated schematically in Fig.\,\ref{fig7}b for $\Delta n$\,=\,0.1. In a well developed phase separation regime we arrive at a system of nearly parallel CO and SF domains separated by supersolid domain walls. However, the regular domain structure becomes more and more unstable the nearer we get to the CO-SF phase transition point. Near $\Delta n\,{\approx}\Delta n_{cr}$ the CO "remnants" do collapse to a 0D skyrmion-like topological defect obviously related with 2$\pi$ domain wall separating the 1D phase domains. This point defect survives up to a maximal doping. Interestingly, such evolution cannot be replicated under SF-CO transition with lowering the doping. The initial stripe structure does not restore, instead, we arrive at creation of an unconventional skyrmion-like cylindrical defect in the CO matrix that does collapse with approaching to the half-filling.

We have also performed computer modeling of the CO-SF phase transition in 2D hc-boson system with a strong Ising anisotropy $V=9t$.
At variance with the previous situation of moderate anisotropy the thermalization procedure for strong anisotropy results in emergence of a well developed rigid domain structure with 180$^{\circ}$ domain walls whose center is characterized by a large nonzero superfluid order parameter which is suppressed  as one runs deep into the CO domain thus suggesting the presence of a filamentary superfluidity nucleated at the antiphase domain walls, that is a gossamer superfluidity, or superfluidity existing in the presence of the "insulating"\, charge order. It is worth noting that the wall width increases under rise of the boson transfer integral.

Weak deviation away from half-filling ($\Delta n\leq$\,0.01) does not give rise to visible modification of the domain structure (see Fig.\,\ref{fig7}c) because the doped bosons do localize in the center of the narrow domain walls breaking the filamentary superfluidity without any visible transformation of the domains. However, further rise of the domain wall loading results in their puzzling transformation in a very small doping range. Tiny amounts of excess bosons suffice to destroy regular narrow domain walls. At the very beginning, the overloading is accompanied by formation of different unstable in-wall structures, in particular, ladder-like patterns formed by bosonic dimers. However, then the "completely filled"\, domain wall becomes regularly broadened because extra bosons prefer to occupy  delocalized states beyond the wall center thus forming a rather extended shell with an inhomogeneous distribution of the SF and SS order parameters. However, the regular wall shape breaks under further doping, these are nonuniformly "swelled"\,  with  the emergence and rise of  widenings.  Step by step these widenings and "blobs"\, nucleated inside domains  spread until these cover all the lattice (see Fig.\,\ref{fig7}c). The Fig.\,\ref{fig7}c does well illustrate a peculiar "memory"\, effect: the CO domain topology survives up to very high doping though the domain wall structure changes significantly.

\section{Conclusion}

We addressed a minimal toy model to describe the charge degree of freedom in the CuO$_2$ planes with the on-site Hilbert space reduced to only a charge triplet of the three effective valence centers (nominally Cu$^{1+;2+;3+}$), and made use of the S=1 pseudospin formalism. It does introduce the on-site mixed valence quantum superpositions,  that, at variance with classical spins or quantum s=1/2 spins,  should  be described by two classical vectors. The formalism provides an unified standpoint for classification of the "myriad"\, of  electronic charge phases in cuprates and their evolution under a nonisovalent doping. Despite its simplicity the S=1 formalism  is shown to constitute a powerful method to describe and study complex phenomena in parent and  doped cuprates, in particular , a comprehensive description of the correlated one- and two-particle transport, coexistence of $p$- and $n$-type carriers, electron-hole asymmetry,  anticorrelation of conventional spin and superconducting order parameters. Concept of the electron and hole centers, differing by a composite (two electrons/holes) local boson, is believed to explain central points of the cuprate puzzles, in particular, the HTSC itself as a condensation of composite bosons and the pseudogap phenomena to be result of the charge order. Our scenario points namely to a charge bosonic degree of freedom engaged by a strong electron-lattice polarization to be responsible for the high-T$_c$ effect in cuprates while the spin degree of freedom does compete with it to reduce high-T$_c$'s.
 The 2D S=1 pseudospin system is prone to a creation of different topological structures which form topologically protected inhomogeneous distributions of the eight local S=1 pseudospin order parameters including charge density and superfluid order parameters. Pseudospin formalism with the two-vector geometrical representation of the on-site states is shown to be the most powerful technique to describe topological structures. We presented a short overview of different localized topological structures which are typical for the S=1 (pseudo)spin systems, in particular, localized topological excitations in “negative-$U$” model which is equivalent to s=1/2 pseudospin system. We argue that even the parent insulating cuprates may be unstable with regard to nucleation of topological defects such as “strange“\, antiphase domain walls or unconventional  localized single- or multi-center skyrmion-like objects with filamentary or ring-shaped superfluid regions. Puzzlingly these  topological structures with complex distribution of the off-diagonal order parameters can be invisible for X-rays due to the same uniform distribution of the mean on-site charge density as in the bare parent monovalent (insulating) phase. Making use of the computer simulation we have demonstrated evolution of different starting charge ordered phases in a model cuprate with large “negative-$U$”   under deviation from half-filling and a step-by-step transformation into superfluid phase. 

In summary, the S=1 pseudospin formalism  is believed to provide a conceptual framework for an in-depth understanding and a novel starting point for analytical and computational studies of high-T$_c$ superconductivity and other puzzles in cuprates.


\begin{acknowledgements}
One of the authors (ASM) would like to thank A. Bianconi, R. Micnas, A. Menushenkov, and S.-L. Drechsler for helpful discussions.
The research was supported by the  Ministry of Education and Science of the Russian Federation, project № FEUZ-2020-0054.
\end{acknowledgements}


\end{document}